\documentclass{elsart}

\usepackage{amssymb}
\usepackage{mathrsfs}
\usepackage{bbm}
\usepackage{graphicx}
\usepackage{floatflt}
\input xy
\xyoption{all}

\def\vecteur#1{#1}
\def\tensdeux#1{\mbox{\boldmath{$#1$}}}
\def\tinytensdeux#1{\mbox{\boldmath{$\scriptstyle{#1}$}}}

\def\tensquatre#1{\mbox{\boldmath{$\mathbbm{#1}$}}}
\def\tinytensquatre#1{\mbox{\boldmath{$\scriptstyle{\mathbbm{#1}}$}}}
\def\sup#1{\mathord{\vtop{\ialign{##\crcr
   $\hfil\mbox{sup}\hfil$\crcr\noalign{\kern1.5pt\nointerlineskip}
   $\hfil\scriptstyle{#1}\hfil$\crcr\noalign{\kern1.5pt}}}}}
\def\inf#1{\mathord{\vtop{\ialign{##\crcr
   $\hfil\mbox{inf}\hfil$\crcr\noalign{\kern1.5pt\nointerlineskip}
   $\hfil\scriptstyle{#1}\hfil$\crcr\noalign{\kern1.5pt}}}}}
\def\lim#1{\mathord{\vtop{\ialign{##\crcr
   $\hfil\mbox{lim}\hfil$\crcr\noalign{\kern1.5pt\nointerlineskip}
   $\hfil\scriptstyle{#1}\hfil$\crcr\noalign{\kern1.5pt}}}}}
\def\intersec#1{\mathord{\vtop{\ialign{##\crcr
   $\hfil\bigcap\hfil$\crcr\noalign{\kern1.5pt\nointerlineskip}
   $\hfil\scriptstyle{#1}\hfil$\crcr\noalign{\kern1.5pt}}}}}
\def\sous#1#2{\mathord{\vtop{\ialign{##\crcr
   $\hfil\displaystyle{#1}\hfil$\crcr\noalign{\kern1.5pt\nointerlineskip}
   $\hfil\scriptstyle{#2}\hfil$\crcr\noalign{\kern1.5pt}}}}}
\def\point#1{\stackrel{\cdot}{#1}}
\def\trace#1{\mbox{tr}(#1)}

\def\nomp#1{#1}

\def\nomc#1{\textit{#1}}
\def\bib#1#2#3{#1~: \textit{#2}. #3.}

\newcommand{\quadcontract}{\raisebox{3.8pt}{$\mathord{\vtop{\ialign{##\crcr
   $\hfil\ : \hfil$\crcr\noalign{\kern1.9pt\nointerlineskip}
   $\hfil\ : \hfil$\crcr\noalign{\kern1.9pt}}}}$}}
\newcommand{\tinyquadcontract}{\raisebox{2.2pt}{$\mathord{\vtop{\ialign{##\crcr
   $\hfil\ : \hfil$\crcr\noalign{\kern1.1pt\nointerlineskip}
   $\hfil\ : \hfil$\crcr\noalign{\kern1.1pt}}}}$}}
\newcommand{\Chi}{\raisebox{2.2pt}{$\mathord{\vtop{\ialign{##\crcr
   $ \chi $\crcr\noalign{\kern1.1pt\nointerlineskip}}}}$}}
\newcommand{\tinyChi}{\raisebox{1.4pt}{$\mathord{\vtop{\ialign{##\crcr
   $ \scriptstyle{\chi} $\crcr\noalign{\kern1.1pt\nointerlineskip}}}}$}}
\newcommand{\reel}{\mathbbm{R}}

\newtheorem{proposition}{Proposition}

\begin{document}

\begin{frontmatter}



\title{Coherent thermodynamical modelling of geomaterial reinforced by wire}


\author[lmgc]{R. Laniel\corauthref{cor_lan}}\ead{romain.laniel@lmgc.univ-montp2.fr}, 
\author[lmgc]{P. Alart}, 
%\corauthref{cor_ala}}\ead{pierre.alart@lmgc.univ-montp2.fr}, 
\author[lmgc]{S. Pagano}
%\ead{stephane.pagano@lmgc.univ-montp2.fr}
\address[lmgc]{LMGC, UMR CNRS 5508, Universit\'e Montpellier II, CC 048 Place Eug\`ene Bataillon, 34095 Montpellier cedex 5, France}
\corauth[cor_lan]{Corresponding author. Phone : +33 4 67 14 45 37}
%\corauth[cor_ala]{Corresponding author. Phone : +33 4 67 14 39 89 ; fax : +33 4 67 14 39 23}
%==============================================================================
%
%  Abstract and Key words
%
%==============================================================================
\begin{abstract}
% Text of abstract
The \nomc{TexSol} is a composite geomaterial : a sand matrix and a wire network reinforcement. For small strains a thermodynamical continuous model of the \nomc{TexSol} including the unilaterality of the wire network is postulated. This model is described by two potentials which depend on some internal variables and a state variable either strain or stress tensor (the choice of this last one gives two different ways of identification). The \nomc{TexSol} continuous model is implemented in a finite element code to recover the mechanical behaviour given by discrete elements numerical experiments.
\end{abstract}
\end{frontmatter}
% main text
%==============================================================================
%
%  1. Motivations
%
%==============================================================================
\section{Motivations}\label{sec:motivations}
%==============================================================================
%
%    1.1. What is the TexSol ?
%
%==============================================================================
\subsection{What is the \nomc{TexSol} ?}\label{ssc:what_is_texsol}
The civil pieces of work need planed stable floor. The environment configuration often forces civil engineers to raise huge embankments. Moreover, it can be interesting to reinforce them to assure a better embankment mechanical behaviour. A lot of different solutions can be used to reinforce soil but, in this paper, we focus our attention to the \nomc{TexSol} process.

The \nomc{TexSol}, created in 1984 by \nomp{Leflaive} \nomp{Khay} and \nomp{Blivet} from LCPC (Laboratoire Central des Ponts et Chauss\'ees) \cite{bib:leflaive_khay_blivet}, is a heterogeneous material by mixing sand and wire network. This reinforced material has a better mechanical resistance than the sand without wire. Of course, the \nomc{TexSol} behaviour depends on sand and wire parameters and its frictional angle can be larger than the sand one from $0^{\circ}$ to $10^{\circ}$ \cite{bib:khay_gigan}. The wire is described by its linear density with a dtex unit ($1 \mbox{ dtex} = 0,1 \mbox{ g}.\mbox{km}^{-1}$), its ponderal content and its stiffness. Classically, the wire density in a \nomc{TexSol} sample is included between $100 \mbox{ km}.\mbox{m}^{-3}$ and $200 \mbox{ km}.\mbox{m}^{-3}$.

\begin{figure}[htbp]
\begin{center}
\begin{tabular}{cc}
\includegraphics[width=5cm]{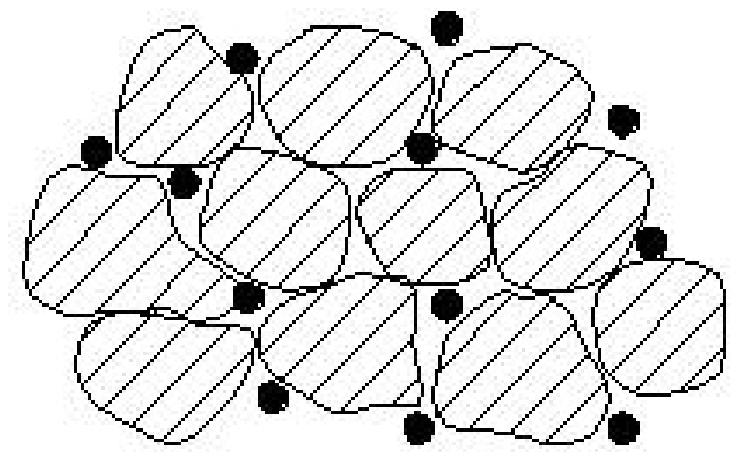} & \includegraphics[width=5cm]{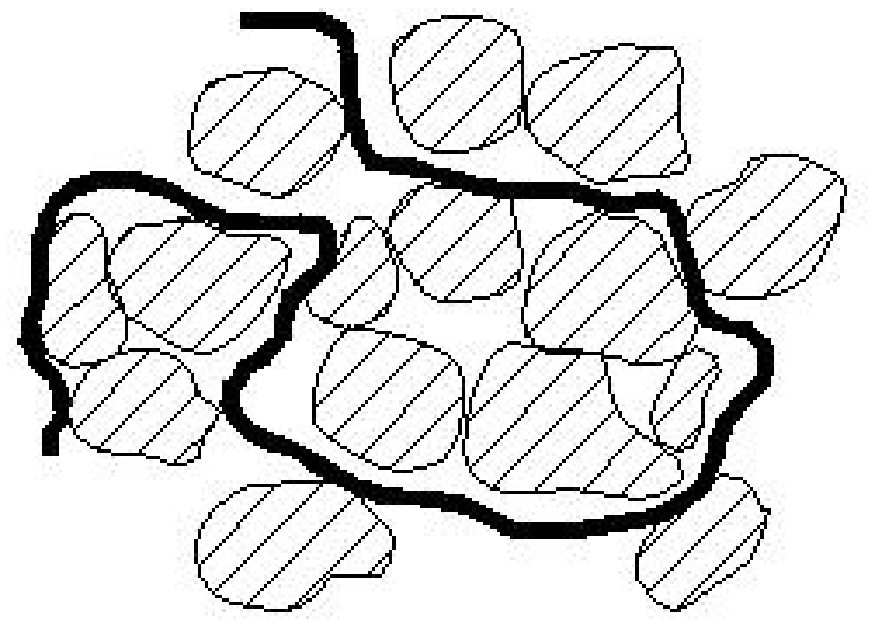}
\end{tabular}
\end{center}
\caption{Schematic \nomc{TexSol} sections}
\label{fig:schem_tex}
\end{figure}

To make a \nomc{TexSol} bank, a machine named \textit{``Texsoleuse''} is used. It works on throwing sand and, in the same time, injecting wire. The wire is deposed on the free plane of the sand with a random orientation. This machine carries out several passes to raise the bank. The figure~\ref{fig:schem_tex} is the \nomc{TexSol} microstructure representation.
In the literature, we find two different continuous modellings. The model suggested in \cite{bib:fremond} is non local and includes remote interactions (corresponding to the wire effects) but also needs an identification of their parameters using macroscopic experiments. \nomp{Villard} proposes a simpler local model in \cite{bib:villard}. This one couples a standard model of sand and an equivalent unilateral elastic stiffness contribution corresponding to the wire network. This last contribution is activated only on the traction directions because of the unilateral behaviour of wire. Our first work is to clearly define thermo-dynamical potentials of the \nomp{Villard} local model with both stress and strain formulations to identify the best-adapted one.
%==============================================================================
%
%    1.2. Assumptions of the continuous local model
%
%==============================================================================
\subsection{Assumptions of the continuous local model}\label{ssc:assumptions}
To couple the elastic plastic model of the sand and the unilateral elastic model of the wire network, we have to consider some mechanical assumptions, which may be backed up by numerical experiments performed with a discrete elements software \cite{{bib:dubois_jean},{bib:moreau_bis}}.
%==============================================================================
%
%      1.2.1. Stress additivity assumption
%
%==============================================================================
\subsubsection{Stress additivity assumption}\label{sss:stress_add}
In this paper, the stress additivity assumption of the sand and the wire network is assumed. Then we write,
\begin{equation}
\label{eqn:stress_add}
\tensdeux{\sigma}_s + \tensdeux{\sigma}_w = \tensdeux{\sigma} \mbox{ ,}
\end{equation}
where $\tensdeux{\sigma}_s$, $\tensdeux{\sigma}_w$ and $\tensdeux{\sigma}$ are respectively the stress second order tensor of the sand, the wire network and the \nomc{TexSol}.

This assumption seems to be coherent with the \nomc{TexSol} quasi-static behaviour. We can get a good approximation of the stress tensor in numerical simulation of 2D granular matter \cite{bib:mouraille} using the \nomp{Weber} stress tensor \cite{bib:cambou_jean}. This tensor may be non symmetrical if inertial effects are not negligeable. For quasi-static processes this discrete tensor is a good candidate to represent a continuous stress tensor. Moreover we can define such a tensor grain by grain with the \nomp{Moreau} approach \cite{bib:moreau_bis}. In this way a wire network stress and a sand stress may be computed, to recover by addition the full \nomc{TexSol} stress (in the simulation, the wire is modeled by a chain of beads with unilateral interactions \cite{bib:laniel_mouraille_pagano_dubois_alart}). On a biaxial crushing test we verify the symmetry property even for large deformation as long as the process remains slow.
\begin{figure}[htbp]
\begin{center}
\vspace{1.5cm}
\includegraphics[width=12cm]{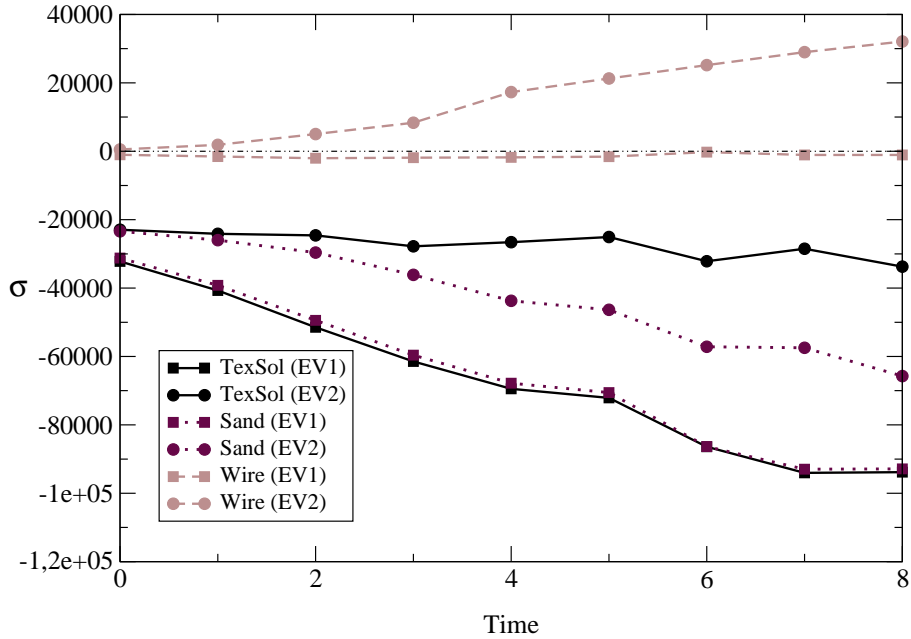}
\vspace{0.5cm}
\end{center}
\caption{Stress eigen values evolution in the \nomc{TexSol}}
\label{fig:eigen_value}
\end{figure}

The eigenvalues $(\mbox{EV}1,\mbox{EV}2)$ are computed and the contribution of each component of the \nomc{TexSol} are underlined in the figure~\ref{fig:eigen_value} where the wire network is only in a tensile state~; in the two eigen directions the sand is in compression. This may be also observed on the distribution of force network in the granular sample.
%==============================================================================
%
%      1.2.2. Non sliding assumption
%
%==============================================================================
\subsubsection{Non sliding assumption}\label{sss:non_sliding}
This second assumption is not as evident as the previous one. Although micro slidings occur between sand grains and wire, we assume that at the macroscopic level of the continuum model, the sand network does not slip through the wire network. This assumption can be translated by the equality of the three strain rates,
\begin{equation}
\label{eqn:non_sliding}
\tensdeux{\point{\varepsilon}}_s=\tensdeux{\point{\varepsilon}}_w=\tensdeux{\point{\varepsilon}} \mbox{ ,}
\end{equation}
where $\tensdeux{\varepsilon}_s$, $\tensdeux{\varepsilon}_w$ and $\tensdeux{\varepsilon}$ are respectively the strain second order tensor of the sand, the wire network and the \nomc{TexSol}.

We have to be very careful with such a condition and define some validity domains for it. Indeed the limits of this assumption are difficult to quantify and we will restrict the validation of the following continuum model to small strains.
%==============================================================================
%
%    1.3. Role of the wire unilaterality
%
%==============================================================================
\subsection{Role of the wire unilaterality}\label{ssc:unilaterality}
The wire network contributes to the tensile srtength of the composite material but not to the compression one (cf. figure~\ref{fig:eigen_value}). To model such a behaviour at the macroscopic scale, it is convenient to introduce a unilateral condition in the behaviour law of the wire network. This unilaterality accounts for two microscopic phenomena. The first one is the lack of bending strength of the wire network viewed as a piece of cotton. The second one is the local buckling of short segments. The first aspect is not explicitely taken into account by a unilateral condition at the microscopic scale in our discrete numerical simulation since the chain of beads has no bending strength. The second aspect may be enforced by introducing a unilateral interaction law between two successive beads.
\begin{figure}[htbp]
\begin{center}
\includegraphics[width=12cm]{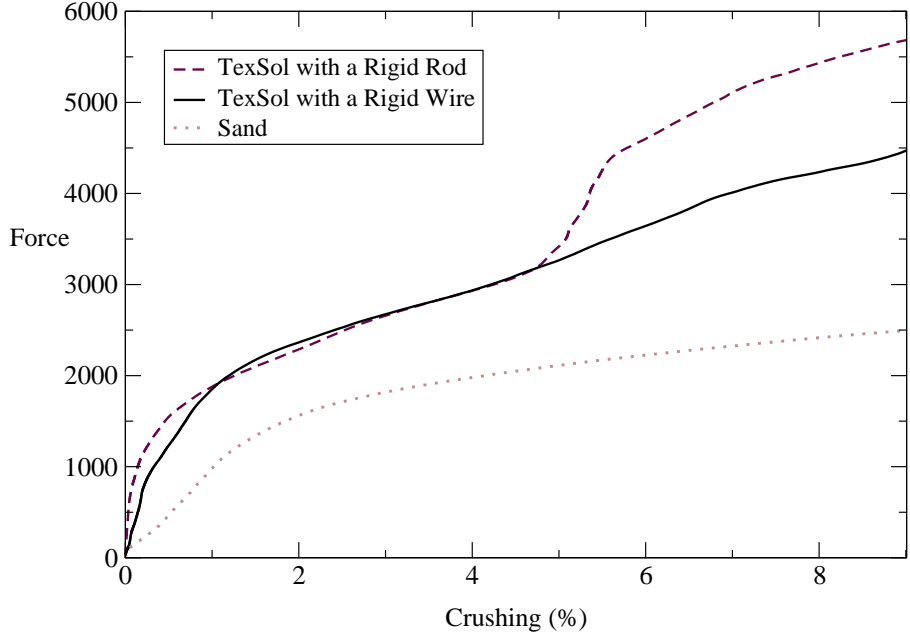}
\end{center}
\caption{Force / Crushing response with different models}
\label{fig:force_crushing}
\end{figure}
\begin{figure}[htbp]
\begin{center}
\includegraphics[width=13cm]{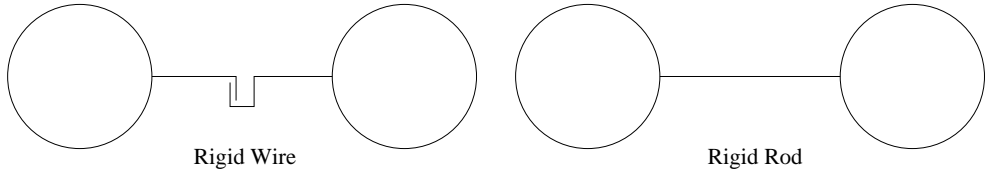}
\end{center}
\caption{Schematic discrete laws}
\label{fig:discrete_laws}
\end{figure}
Such an interaction models an elemental wire between two beads~: we denote by ``rigid wire''; if not we speak about ``rigid rod'' (cf. figure \ref{fig:discrete_laws}) for bilateral law between beads. The figure \ref{fig:force_crushing} illustrates the difference of global behaviour between both simulation for crushing biaxial test. Until $6$ percents of deformation, the responses are almost identical. For larger deformation the ``rigid rod'' model leads to a rough increase of rigidity due to the appearance of compression columns in the wire. Such a phenomenon seems not very realistic and is probably issued from a scale effect since the numerical sample is not representative enough of the material. In particular the model of wire with a chain of beads generates non realistic wedges of beads by sand grains.
%==============================================================================
%
%    1.4. Why a strain and stress formulation ?
%
%==============================================================================
\subsection{Why a strain and stress formulation ?}\label{ssc:strain_stress_formula}
In this paper, we propose to carry out a thermodynamical study with both strain and stress formulations. The interest of this work is in the identification possibilities of potentials parameters. Indeed, an experimentator making some tests on a sample has only access to the \textit{global} strain. Our numerical investigations allow us to have access to finer data such as the \textit{local} stress field throughout the sample. Moreover the \textit{global} stress tensor over the sample can be deduced by an average.

The post processing of numerical experiments mentioned in \S\ref{sss:stress_add} provides precise informations on the stress fields, in the sand and in the wire network. The stress ``unilaterality'' in the wire is clearly established in the figure \ref{fig:eigen_value}. This observation could lead us to favour a stress formulation. But the finite element softwares are essentially developed using a strain formulation. Consequently we propose, in the following study, strain formulations easily implementable. Dual stress formulations are provided when they can be analytically deduced by the \nomp{Legendre~--~Fenchel} transformation.
%==============================================================================
%
%  2. A general thermodynamical framework
%
%==============================================================================
\section{A general thermodynamical framework}\label{sec:thermo}
In this part, we define potentials written with different state variables. These potentials have to check the \nomp{Clausius--Duhem} inequality to be thermodynamically admissible.
%==============================================================================
%
%    2.1. Strain versus stress approach in thermodynamics
%
%==============================================================================
\subsection{Strain versus stress approach in thermodynamics}\label{ssc:stress_strain}
This work must be as exhaustive as possible, while passing from unspecified state variables to its dual. We thus use the \nomp{Legendre--Fenchel} transformation \cite{bib:moreau}, to carry out our study with both strain and stress formulations. Let us write the \nomp{Clausius--Duhem} inequality where $u$ is the internal energy, $s$ the entropy, $\vecteur{q}$ the heat flow and $T$ the temperature,
\begin{equation}
\label{eqn:clausius}
\tensdeux{\sigma} : \point{\tensdeux{\varepsilon}} - \rho\left(\point{u} - T\point{s}\right) - \frac{1}{T}\vecteur{q} . \vecteur{\bigtriangledown}T \geq 0 \mbox{ .}
\end{equation}
The intrinsic dissipation depends on a state variable $\tensdeux{X}$ (or its dual $\tensdeux{X}^*$), some internal variables $\alpha = \left\{\alpha_1,\cdots,\alpha_p\right\}$ (each internal variable can be scalar, vectorial or tensorial) and the temperature $T$. It can also be expressed with the free energy $\psi$ or its \nomp{Legendre--Fenchel} transformation $\psi^*$ with respect to the state variable $\tensdeux{X}$,
$$
\begin{array}{c}
 \\
\mbox{and  } \\
 \\
\end{array}
\begin{array}{l}
\begin{array}{l}
\psi(\tensdeux{X}, \alpha, T) = u(\tensdeux{X}, \alpha, T) - T s(\tensdeux{X}, \alpha, T) \\
\end{array} \\
\begin{array}{ll}
\psi^*(\tensdeux{X}^*, \alpha, T) & = \sup{\tinytensdeux{\overline{X}}}\left\{\tensdeux{\overline{X}} : \tensdeux{X}^* - \psi(\tensdeux{\overline{X}}, \alpha, T)\right\} \\
 & = \tensdeux{X} : \tensdeux{X}^* - \psi(\tensdeux{X}, \alpha, T) \mbox{ ,} \\
\end{array} \\
\end{array}
$$
where $\tensdeux{X}$ is the argument of the supremum. Considering either $\tensdeux{X}$ or $\tensdeux{X}^*$, we find two expressions of the \nomp{Clausius--Duhem} inequality,
\begin{equation}
\label{eqn:clausius_x}
\begin{array}{c}
\tensdeux{\sigma} : \point{\tensdeux{\varepsilon}} - \rho\left[\frac{\partial \psi}{\partial \tinytensdeux{X}} : \point{\tensdeux{X}} + \left(s + \frac{\partial \psi}{\partial T}\right)\point{T} + \frac{\partial \psi}{\partial \alpha_m}\point{\alpha}_m\right] - \frac{\vecteur{q}}{T} . \vecteur{\bigtriangledown}T \geq 0
\end{array}\mbox{ ,}
\end{equation}
\begin{equation}
\label{eqn:clausius_x_star}
\begin{array}{rcl}
\tensdeux{\sigma} : \point{\tensdeux{\varepsilon}} - \rho\left[\point{\tensdeux{X}} : \tensdeux{X}^* + \left(\tensdeux{X} - \frac{\partial \psi^*}{\partial \tinytensdeux{X}^{\scriptscriptstyle{*}}}\right) : \point{\tensdeux{X}}^* + \left(s - \frac{\partial \psi^*}{\partial T}\right)\point{T} - \frac{\partial \psi^*}{\partial \alpha_m}\point{\alpha}_m\right] & & \\
- \frac{\vecteur{q}}{T} . \vecteur{\bigtriangledown}T & \geq & 0\mbox{ .}
\end{array}
\end{equation}
Using the \nomp{Helmholtz} postulate (this last one can be applied with the generalized standard materials assumption \cite{bib:halphen_nguyen}) and the previous definitions, we are now able to deduce the state laws,
\begin{equation}
\label{eqn:state_law}
\begin{array}{cccc}
\mbox{Primal state laws} & \qquad & \mbox{Dual state laws} & \\
 & & & \\
\tensdeux{X}^* \in \partial_{\tinytensdeux{X}}\psi(\tensdeux{X}, \alpha, T) & & \tensdeux{X} \in \partial_{\tinytensdeux{X}^*}\psi^*(\tensdeux{X}^*, \alpha, T) & \\
-s \in \partial_T\psi(\tensdeux{X}, \alpha, T) & & s \in \partial_T\psi^*(\tensdeux{X}^*, \alpha, T) & \\
-\frac{1}{\rho}A_m \in \partial_{\alpha_m}\psi(\tensdeux{X}, \alpha, T) & & \frac{1}{\rho}A_m \in \partial_{\alpha_m}\psi^*(\tensdeux{X}^*, \alpha, T) & \mbox{,}\\
\end{array}
\end{equation}
where $A_m$ is the thermodynamical force associated with $\alpha_m$. Formally we use subdifferentials instead of derivatives. If convexity is not required, previous relations still hold using the \nomp{Clarke} subdifferential \cite{bib:clarke}. Then the primal and dual forms are not necessary equivalent. In the general case, the \nomp{Clausius--Duhem} inequality (\ref{eqn:clausius_x}) or (\ref{eqn:clausius_x_star}) can be reduced to a dot product of a vector flow and a vector force,
\begin{equation}
\label{eqn:clausius_vec}
{\underbrace{\left[\begin{array}{c}\widehat{\tensdeux{\sigma}}\\A\\\vecteur{\bigtriangledown}T\end{array}\right]}_{\mbox{force}}}^T.\underbrace{\left[\begin{array}{c}\point{\widehat{\tensdeux{\varepsilon}}}\\\point{\alpha}\\-\frac{\vecteur{q}}{T}\end{array}\right]}_{\mbox{flow}} \geq 0 \qquad \mbox{where } \left\{ \begin{array}{rcl}\point{\widehat{\tensdeux{\varepsilon}}}&=&\point{\tensdeux{\varepsilon}}\mbox{ or }\point{\tensdeux{\varepsilon}}^{ir}\\\widehat{\tensdeux{\sigma}}&=&\tensdeux{\sigma}^{ir}\mbox{ or }\tensdeux{\sigma}\end{array} \right.\mbox{ .}
\end{equation}
The flow variables have to be related by evolution laws to the force variables. To verify the inequality (\ref{eqn:clausius_vec}) some assumptions may be added to these relations. It is convenient to introduce a dissipation potential $\varphi$ from which the evolution laws are derived. By duality a force function $\varphi^*$ is automatically defined using the \nomp{Legendre--Fenchel} transformation,
\begin{equation}
\label{eqn:complementary_law}
\begin{array}{cccc}
\mbox{Primal complementary laws} & \qquad & \mbox{Dual complementary laws} & \\
 & & & \\
\widehat{\tensdeux{\sigma}} \in \partial_{\point{\widehat{\tinytensdeux{\varepsilon}}}}\varphi(\point{\widehat{\tensdeux{\varepsilon}}},\point{\alpha},-\frac{\vecteur{q}}{T}) & & \point{\widehat{\tensdeux{\varepsilon}}} \in \partial_{\widehat{\tinytensdeux{\sigma}}}\varphi^*(\widehat{\tensdeux{\sigma}},\point{\alpha},-\frac{\vecteur{q}}{T}) & \\
A \in \partial_{\point{\alpha}}\varphi(\point{\widehat{\tensdeux{\varepsilon}}},\point{\alpha},-\frac{\vecteur{q}}{T}) & & -A \in \partial_{\point{\alpha}}\varphi^*(\widehat{\tensdeux{\sigma}},\point{\alpha},-\frac{\vecteur{q}}{T}) & \\
\vecteur{\bigtriangledown}T \in \partial_{\left(-\frac{\vecteur{q}}{T}\right)}\varphi(\point{\widehat{\tensdeux{\varepsilon}}},\point{\alpha},-\frac{\vecteur{q}}{T}) & & -\vecteur{\bigtriangledown}T \in \partial_{\left(-\frac{\vecteur{q}}{T}\right)}\varphi^*(\widehat{\tensdeux{\sigma}},\point{\alpha},-\frac{\vecteur{q}}{T}) & \mbox{.}\\
\end{array}
\end{equation}
To verify the \nomp{Clausius--Duhem} inequality, some assumptions on the dissipation potential are necessary. For simplicity we consider now an isothermal process. The left-hand side of the inequality is reduced to,
$$
\widehat{\tensdeux{\sigma}}:\point{\widehat{\tensdeux{\varepsilon}}} + A\point{\alpha} \; = \frac{\partial \varphi}{\partial \point{\widehat{\tensdeux{\varepsilon}}}}:\point{\widehat{\tensdeux{\varepsilon}}} + \frac{\partial \varphi}{\partial \point{\alpha}}\point{\alpha} \; = \left\langle\partial \varphi(\point{\widehat{\tensdeux{\varepsilon}}}, \point{\alpha}), \left(\point{\widehat{\tensdeux{\varepsilon}}}, \point{\alpha}\right)\right\rangle
$$
and the primal state laws are summarized in $\left(\widehat{\tensdeux{\sigma}}, A\right) \in \partial \varphi(\point{\widehat{\tensdeux{\varepsilon}}}, \point{\alpha})$. $\varphi$ being a \textit{separately convex function}, with a convex analysis characterisation of the subdifferential we write,
$$
\forall \left(\tensdeux{x}, y\right) \quad \varphi(\point{\widehat{\tensdeux{\varepsilon}}}, \point{\alpha}) - \varphi(\tensdeux{x}, y) \leq \left\langle\left(\point{\widehat{\tensdeux{\varepsilon}}}, \point{\alpha}\right) - \left(\tensdeux{x}, y\right), \left(\widehat{\tensdeux{\sigma}}, A\right)\right\rangle\mbox{ .}
$$
Moreover, if \textit{$\varphi$ is minimum in $\left(\tensdeux{0}, 0\right)$}, the \nomp{Clausius--Duhem} inequality is then verified \cite{bib:suquet},
$$
\widehat{\tensdeux{\sigma}}:\point{\widehat{\tensdeux{\varepsilon}}} + A\point{\alpha} \; \geq \varphi(\point{\widehat{\tensdeux{\varepsilon}}}, \point{\alpha}) - \varphi(\tensdeux{0}, 0) \geq 0 \mbox{ .}
$$
Similar properties are required for $\varphi^*$ to recover the \nomp{Clausius--Duhem} inequality. Generally we distinguish the reversible and irreversible parts of the transformation. We thus postulate an additive decomposition for both reversible and irreversible parts of the strain tensor $\tensdeux{\varepsilon} = \tensdeux{\varepsilon}^r + \tensdeux{\varepsilon}^{ir}$ and the stress tensor $\tensdeux{\sigma} = \tensdeux{\sigma}^r + \tensdeux{\sigma}^{ir}$. The reversible / irreversible splitting of $\tensdeux{\sigma}$ is less classical. To illustrate its interest, remark that eventual residual stresses may be accounted for in the irreversible part.

At this stage we have to choose the external state variable $\tensdeux{X}$ for the strain formulation and consequently $\tensdeux{X}^*$ for the stress formulation. It is usual to consider for $\tensdeux{X}$ the total strain tensor $\tensdeux{\varepsilon}$. By the way the reversible stress $\tensdeux{\sigma}^r$ appears in the state law and becomes the state variable in the dual stress formulation. But we can also use the reversible strain part $\tensdeux{\varepsilon}^r$ and deduce the full stress tensor $\tensdeux{\sigma}$ as the dual state variable (cf. table \ref{tab:resume_thermo}).
\begin{table}[htbp]
\begin{center}
\scriptsize{
\begin{tabular}{||p{3cm}|p{3cm}||p{3cm}|p{3cm}||}
\hline\hline
\multicolumn{2}{||l||}{\textit{\textbf{State variable~: $\tensdeux{\varepsilon}^r$}}}
&
\multicolumn{2}{|l||}{\textit{\textbf{State variable~: $\tensdeux{\sigma}$}}}
\\
\multicolumn{2}{||c||}{$\tensdeux{\sigma} : \point{\tensdeux{\varepsilon}}^{ir} + A_m\point{\alpha}_m - \frac{\vecteur{q}}{T}\vecteur{\bigtriangledown}T \geq 0$}
&
\multicolumn{2}{|c||}{$\point{\tensdeux{\varepsilon}}^{ir} : \tensdeux{\sigma} + A_m\point{\alpha}_m - \frac{\vecteur{q}}{T}\vecteur{\bigtriangledown}T \geq 0$}
\\\hline
free energy~: $\psi$
&
Dissipation potential~: $\varphi$
&
Free enthalpy~: $\psi^*$
&
Force function~: $\varphi^*$
\\\hline
\multicolumn{1}{||c|}{$\frac{1}{\rho}\tensdeux{\sigma} \in \partial_{\tinytensdeux{\varepsilon}^r}\psi$}
&
\multicolumn{1}{|c||}{$\tensdeux{\sigma} \in \partial_{\tinytensdeux{\point{\varepsilon}}^{ir}} \varphi$}
&
\multicolumn{1}{|c|}{$\frac{1}{\rho}\tensdeux{\varepsilon}^r \in \partial_{\tinytensdeux{\sigma}} \psi^*$}
&
\multicolumn{1}{|c||}{$\tensdeux{\point{\varepsilon}}^{ir} \in \partial_{\tinytensdeux{\sigma}} \varphi^*$}
\\
\multicolumn{1}{||c|}{$- s \in \partial_T \psi$}
&
\multicolumn{1}{|c||}{$A_m \in \partial_{\point{\alpha}_m}\varphi$}
&
\multicolumn{1}{|c|}{$s \in \partial_T \psi^*$}
&
\multicolumn{1}{|c||}{$- A_m \in \partial_{\point{\alpha}_m}\varphi^*$}
\\
\multicolumn{1}{||c|}{$- \frac{1}{\rho}A_m \in \partial_{\alpha_m}\psi$}
&
\multicolumn{1}{|c||}{$\vecteur{\bigtriangledown}T \in \partial_{\scriptscriptstyle{\left(- \frac{\vecteur{q}}{T}\right)}} \varphi$}
&
\multicolumn{1}{|c|}{$\frac{1}{\rho}A_m \in \partial_{\alpha_m}\psi^*$}
&
\multicolumn{1}{|c||}{$- \vecteur{\bigtriangledown}T \in \partial_{\scriptscriptstyle{\left(- \frac{\vecteur{q}}{T}\right)}} \varphi^*$}
\\\hline\hline
\multicolumn{2}{||l||}{\textit{\textbf{State variable~: $\tensdeux{\varepsilon}$}}}
&
\multicolumn{2}{|l||}{\textit{\textbf{State variable~: $\tensdeux{\sigma}^r$}}}
\\
\multicolumn{2}{||c||}{$\tensdeux{\sigma}^{ir} : \point{\tensdeux{\varepsilon}} + A_m\point{\alpha}_m - \frac{\vecteur{q}}{T}\vecteur{\bigtriangledown}T \geq 0$}
&
\multicolumn{2}{|c||}{$\point{\tensdeux{\varepsilon}} : \tensdeux{\sigma}^{ir} + A_m\point{\alpha}_m - \frac{\vecteur{q}}{T}\vecteur{\bigtriangledown}T \geq 0$}
\\\hline
Free energy~: $\psi$
&
Dissipation potential~: $\varphi$
&
Free enthalpy~: $\psi^*$
&
Force function~: $\varphi^*$
\\\hline
\multicolumn{1}{||c|}{$\frac{1}{\rho}\tensdeux{\sigma}^r \in \partial_{\tinytensdeux{\varepsilon}}\psi$}
&
\multicolumn{1}{|c||}{$\tensdeux{\sigma}^{ir} \in \partial_{\point{\tinytensdeux{\varepsilon}}}\varphi$}
&
\multicolumn{1}{|c|}{$\frac{1}{\rho}\tensdeux{\varepsilon} \in \partial_{\tinytensdeux{\sigma}^r}\psi^*$}
&
\multicolumn{1}{|c||}{$\point{\tensdeux{\varepsilon}} \in \partial_{\tinytensdeux{\sigma}^{ir}}\varphi^*$}
\\
\multicolumn{1}{||c|}{$- s \in \partial_T \psi$}
&
\multicolumn{1}{|c||}{$A_m \in \partial_{\point{\alpha}_m}\varphi$}
&
\multicolumn{1}{|c|}{$s \in \partial_T \psi^*$}
&
\multicolumn{1}{|c||}{$- A_m \in \partial_{\point{\alpha}_m} \varphi^*$}
\\
\multicolumn{1}{||c|}{$- \frac{1}{\rho}A_m \in \partial_{\alpha_m}\psi$}
&
\multicolumn{1}{|c||}{$\vecteur{\bigtriangledown}T \in \partial_{\scriptscriptstyle{\left(- \frac{\vecteur{q}}{T}\right)}}\varphi$}
&
\multicolumn{1}{|c|}{$\frac{1}{\rho}A_m \in \partial_{\alpha_m} \psi^*$}
&
\multicolumn{1}{|c||}{$- \vecteur{\bigtriangledown}T \in \partial_{\scriptscriptstyle{\left(- \frac{\vecteur{q}}{T}\right)}}\varphi^*$}
\\\hline\hline
\multicolumn{4}{c}{}
\end{tabular}}
\caption{Strain versus stress formulations}
\label{tab:resume_thermo}
\end{center}
\end{table}
The first column expresses the primal model using $\tensdeux{\varepsilon}^r$ or $\tensdeux{\varepsilon}$ as state variable. The second one provides the corresponding dual formulations.
%==============================================================================
%
%    2.2. 1D model of reinforced geomaterial
%
%==============================================================================
\subsection{1D model of reinforced geomaterial}\label{ssc:texsol_1D}
Let us apply previous results to a rheological 1D model of \nomc{TexSol} taking into account the wire unilaterality.
%==============================================================================
%
%      2.2.1. Strain formulation
%
%==============================================================================
\subsubsection{Strain formulation}\label{sss:1D_strain}
\begin{figure}[htbp]
\begin{center}
\includegraphics[width=10cm]{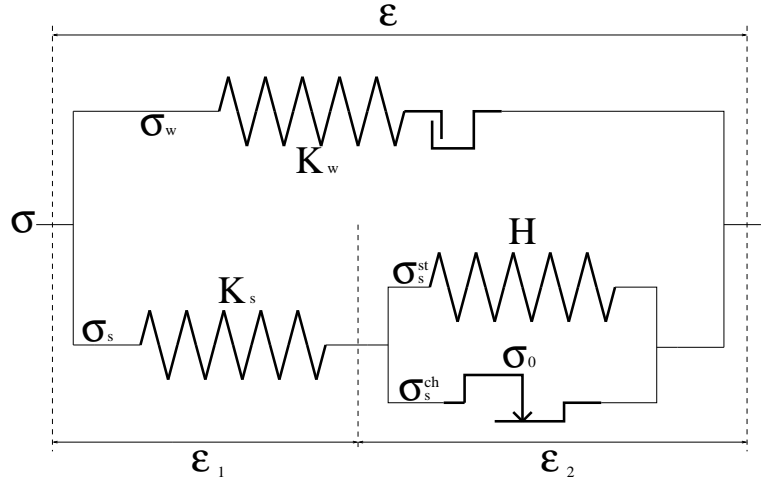}
\end{center}
\caption{Rheological \nomc{TexSol} diagram}
\label{fig:texsol}
\end{figure}

We choose to superpose a classical 1D model of elasto-plasticity with hardening for sand \cite{bib:lemaitre_chaboche} and a 1D unilateral model of elasticity for wire. We thus propose the two potentials $\psi$ (free energy) and $\varphi$ (dissipation potential) depending on the external state variable $\varepsilon$ and on the internal one $\varepsilon_2$ as shown in the figure~\ref{fig:texsol},
\begin{equation}
\label{eqn:psi_tex_1D}
\psi\left(\varepsilon, \varepsilon_2\right) = \left\{\begin{array}{ll} \psi_1(\varepsilon,\varepsilon_2) & \quad \mbox{if }\varepsilon \in C_1 \\ \psi_2(\varepsilon,\varepsilon_2) & \quad \mbox{if }\varepsilon \in C_2 \\ \end{array}\right.\mbox{ ,}
\end{equation}
\begin{equation}
\label{eqn:phi_tex_1D}
\varphi(\point{\varepsilon}, \point{\varepsilon_2}) = \sigma_0\left|\point{\varepsilon_2}\right|\mbox{ ,}
\end{equation}
where $\begin{array}{l}\psi_1(\varepsilon,\varepsilon_2) = \frac{1}{2}K_w\varepsilon^2 + \frac{1}{2}K_s\left(\varepsilon - \varepsilon_2\right)^2 + \frac{1}{2}H{\varepsilon_2}^2 \\ \psi_2(\varepsilon,\varepsilon_2) = \frac{1}{2}K_s\left(\varepsilon - \varepsilon_2\right)^2 + \frac{1}{2}H{\varepsilon_2}^2 \\ \end{array}$, $\begin{array}{l}C_1 = \left\{\varepsilon\in\reel\left|\varepsilon \geq 0\right.\right\} \\ C_2 = \left\{\varepsilon\in\reel\left|\varepsilon \leq 0\right.\right\} \\ \end{array}$ and $\sigma_0$ the stress threshold. According to the table \ref{tab:resume_thermo} the state and complementary laws are derived,
\begin{flushleft}
$\quad \begin{array}{l}
\mbox{State laws} \\
\sigma^r = \frac{\partial \psi}{\partial \varepsilon} = K_w\langle\varepsilon\rangle + K_s\left(\varepsilon - \varepsilon_2\right) \mbox{ , } A = - \frac{\partial \psi}{\partial \varepsilon_2} = K_s\left(\varepsilon - \varepsilon_2\right) - H\varepsilon_2\\
\end{array}$
\end{flushleft}
\begin{flushleft}
$\quad \begin{array}{l}
\mbox{Complementary laws} \\
\sigma^{ir} = \frac{\partial \varphi}{\partial \point{\varepsilon}} = 0 \mbox{ , } A \in \partial_{\point{\varepsilon_2}}\varphi = \left\{\begin{array}{ll} \left\{sign(\point{\varepsilon_2})\sigma_0\right\} & \quad\mbox{if } \point{\varepsilon_2} \in \reel^* \\ \left[- \sigma_0, \sigma_0\right] & \quad\mbox{if } \point{\varepsilon_2} = 0 \end{array}\right. \mbox{ ,}\\
\end{array}$
\end{flushleft}
where $\langle\varepsilon\rangle = \mbox{max}(0,\varepsilon)$, the non negative part of $\tensdeux{\varepsilon}$.
%==============================================================================
%
%      2.2.2. Stress formulation
%
%==============================================================================
\subsubsection{Stress formulation}\label{sss:1D_stress}
To determine the stress formulation, we have to calculate the \nomp{Legendre--Fenchel} transformations of $\psi$ and $\varphi$ which are not always analytically accessible. However we can use the following general result convenient for piecewise smooth functions.
\begin{proposition}
\label{prp:psi_conj_theo}
Consider a non overlapping splitting ${\left(C_i\right)}_{i=1,n}$ of the strain space $\reel^{3 \times 3}$, $\bigcup_{i=1}^n C_i = \reel^{3 \times 3}$, $C_i$ close convex cone with $\mbox{mes}(C_i \cap C_j) = 0$, $i \neq j$. If $\psi(\tensdeux{\varepsilon})$ is piecewise defined by $\psi(\tensdeux{\varepsilon}) = \psi_i(\tensdeux{\varepsilon})$ if $\tensdeux{\varepsilon} \in C_i \mbox{, } i=1,\dots,n$ then 
$$
\psi^*(\tensdeux{\sigma}) = \sup{i}\left\{\left(\psi_i^*\bigtriangledown \textup{I}_{C_i^{\circ}}\right)(\tensdeux{\sigma})\right\}\mbox{ .}
$$
\end{proposition}

\textbf{Proof :} Let recall the definition of inf-convolution of two functions $f$ and $g$ \cite{bib:moreau}, the indicator function of a convex set $A$ and the polar cone $C^{\circ}$ of $C$,
$$
\begin{array}{l}
\circ \quad \left(f\bigtriangledown g\right)(\varepsilon) = \inf{\varepsilon=\varepsilon_1+\varepsilon_2}\{f(\varepsilon_1)+f(\varepsilon_2)\} \\
\circ \quad \textup{I}_A(\varepsilon) = \left\{\begin{array}{cl}0 & \mbox{ if }\varepsilon \in A \\ +\infty & \mbox{ if }\varepsilon \not\in A \\\end{array}\right.\\
\circ \quad C^{\circ} = \left\{\tensdeux{\sigma}\left|\tensdeux{\varepsilon}:\tensdeux{\sigma}\leq 0 \mbox{ , }\forall \tensdeux{\varepsilon}\in C\right.\right\}
\end{array}
$$
According to classical rules of convex analysis,
$$
\begin{array}{rcl}
\psi^*(\tensdeux{\sigma}) & = & \sup{\tinytensdeux{\overline{\varepsilon}}}\left\{\tensdeux{\sigma}:\tensdeux{\overline{\varepsilon}} - \inf{i}\left\{\psi_i(\tensdeux{\overline{\varepsilon}}) + \textup{I}_{C_i}(\tensdeux{\overline{\varepsilon}})\right\}\right\} \\
                          & = & \sup{i}\left\{\sup{\tinytensdeux{\overline{\varepsilon}}}\left\{\tensdeux{\sigma}:\tensdeux{\overline{\varepsilon}} - \psi_i(\tensdeux{\overline{\varepsilon}}) - \textup{I}_{C_i}(\tensdeux{\overline{\varepsilon}})\right\}\right\} \\
                          & = & \sup{i}\left\{\left(\psi_i + \textup{I}_{C_i}\right)^*(\tensdeux{\sigma})\right\} \\
                          & = & \sup{i}\left\{\left(\psi_i^*\bigtriangledown \textup{I}_{C_i^{\circ}}\right)(\tensdeux{\sigma})\right\} \mbox{ .}\\
\end{array}
$$

For the 1D model the splitting into two half spaces is obvious and the analytical forms of conjugate functions from (\ref{eqn:psi_tex_1D}) are reachable,
$$
\begin{array}{ll}
\psi^*_1(\sigma^r,\varepsilon_2) = \frac{\left(\sigma^r + K_s\varepsilon_2\right)^2}{2\left(K_w + K_s\right)} - \frac{\left(K_s + H\right){\varepsilon_2}^2}{2} & \mbox{ ,  } C_1^{\circ} = \left\{\sigma\in\reel\left|\sigma \leq 0\right.\right\} \\
\psi^*_2(\sigma^r,\varepsilon_2) = \frac{\left(\sigma^r + K_s\varepsilon_2\right)^2}{2K_s} - \frac{\left(K_s + H\right){\varepsilon_2}^2}{2} & \mbox{ ,  } C_2^{\circ} = \left\{\sigma\in\reel\left|\sigma \geq 0\right.\right\} \mbox{ .}\\
\end{array}
$$
Using the proposition \ref{prp:psi_conj_theo}, we obtain successively,
$$
\begin{array}{l}
\psi_1^*\bigtriangledown \textup{I}_{C_1^{\circ}}=\left\{\begin{array}{ll} - \frac{\left(K_s + H\right){\varepsilon_2}^2}{2} & \mbox{if } \scriptstyle{\sigma^r + K_s\varepsilon_2 \leq 0} \\ \frac{\left(\sigma^r + K_s\varepsilon_2\right)^2}{2\left(K_w + K_s\right)} - \frac{\left(K_s + H\right){\varepsilon_2}^2}{2} & \mbox{if } \scriptstyle{\sigma^r + K_s\varepsilon_2 \geq 0} \\ \end{array}\right. \\
\psi_2^*\bigtriangledown \textup{I}_{C_2^{\circ}}=\left\{\begin{array}{ll} - \frac{\left(K_s + H\right){\varepsilon_2}^2}{2} & \mbox{if } \scriptstyle{\sigma^r + K_s\varepsilon_2 \geq 0} \\ \frac{\left(\sigma^r + K_s\varepsilon_2\right)^2}{2K_s} - \frac{\left(K_s + H\right){\varepsilon_2}^2}{2} & \mbox{if } \scriptstyle{\sigma^r + K_s\varepsilon_2 \leq 0}\mbox{ .} \\ \end{array}\right. \\
\end{array}
$$
Finally,
\begin{equation}
\label{eqn:psi_star_tex_1D}
\psi^*(\sigma^r,\varepsilon_2) = \left\{\begin{array}{ll} \frac{\left(\sigma^r + K_s\varepsilon_2\right)^2}{2\left(K_w + K_s\right)} - \frac{\left(K_s + H\right){\varepsilon_2}^2}{2} & \mbox{ if } \sigma^r + K_s\varepsilon_2 \geq 0 \\ \frac{\left(\sigma^r + K_s\varepsilon_2\right)^2}{2K_s} - \frac{\left(K_s + H\right){\varepsilon_2}^2}{2} & \mbox{ if } \sigma^r + K_s\varepsilon_2 \leq 0 \\ \end{array}\right. \mbox{ .}\\
\end{equation}
The \nomp{Legendre--Fenchel} transformation of the dissipation potential is computed classically from (\ref{eqn:phi_tex_1D}),
\begin{equation}
\label{eqn:phi_star_tex_1D}
\varphi^*(\sigma^{ir}, \point{\varepsilon_2}) = \textup{I}_{\{0\}}(\sigma^{ir}) - \sigma_0\left|\point{\varepsilon_2}\right| \mbox{ .}
\end{equation}
We implicitely get from the equation (\ref{eqn:phi_star_tex_1D})~: $\sigma^{ir} = 0$. The state and complementary laws in the stress formulation are straightforward derived,
\begin{flushleft}
$\quad \begin{array}{l}
\mbox{State laws} \\
\varepsilon = \frac{\partial \psi^*}{\partial \sigma^r}(\sigma^{ir}, \varepsilon_2) = \left\{\begin{array}{ll}\frac{\sigma^r + K_s\varepsilon_2}{K_w + K_s} & \mbox{\scriptsize{if }} \scriptstyle{\sigma^r + K_s\varepsilon_2 \geq 0} \\ \frac{\sigma^r + K_s\varepsilon_2}{K_s} & \mbox{\scriptsize{if }} \scriptstyle{\sigma^r + K_s\varepsilon_2 \leq 0} \\ \end{array}\right. \\
A = \frac{\partial \psi^*}{\partial \varepsilon_2}(\sigma^{ir}, \varepsilon_2) = \left\{\begin{array}{ll}\scriptstyle{\frac{K_s}{K_w + K_s}\left(\sigma^r + K_s\varepsilon_2\right) - \left(K_s + H\right)\varepsilon_2} & \mbox{\scriptsize{if }} \scriptstyle{\sigma^r + K_s\varepsilon_2 \geq 0} \\ \scriptstyle{\sigma^r + K_s\varepsilon_2 - \left(K_s + H\right)\varepsilon_2} & \mbox{\scriptsize{if }} \scriptstyle{\sigma^r + K_s\varepsilon_2 \leq 0} \\ \end{array}\right. \\
\end{array}$
\end{flushleft}
\begin{flushleft}
$\quad \begin{array}{l}
\mbox{Complementary laws} \\
\point{\varepsilon} \in \partial_{\sigma^{ir}} \varphi^*(\sigma^{ir}, \point{\varepsilon_2}) = \reel \\
A \in - \partial_{\point{\varepsilon_2}} \varphi^*(\sigma^{ir}, \point{\varepsilon_2}) = \left\{\begin{array}{ll} \left\{sign(\point{\varepsilon_2})\sigma_0\right\} & \mbox{si } \point{\varepsilon_2} \in \reel^* \\ \left[- \sigma_0, \sigma_0\right] & \mbox{si } \point{\varepsilon_2} = 0 \end{array}\right. \mbox{ .} \\
\end{array}$
\end{flushleft}
This set of equations is equivalent to the one obtained with the strain formulation \S \ref{sss:1D_strain}.
%==============================================================================
%
%    2.3. Extention to a rigid plastic modelisation
%
%==============================================================================
%\subsection{Extention to a rigid plastic modelisation}\label{ssc:rigid_plastic}
%==============================================================================
%
%  3. Strain and stress approach for 3D models
%
%==============================================================================
\section{Strain and stress approach for 3D models}\label{sec:stress_3D}
The complex microstructure of the \nomc{TexSol} material needs not to neglect the three dimensional effects. To define a 3D model we follow the previous 1D approach superposing a classical elastic plastic behaviour for the sand and a unilateral elastic one for the wire network. Simple and sophisticated unilaterality conditions may be considered leading to different formulations more or less easy to handle in a general primal / dual framework.
%==============================================================================
%
%    3.1. 3D thermodynamical potentials of the sand
%
%==============================================================================
\subsection{3D thermodynamical potentials of the sand}\label{ssc:sand_3D}
First of all, let us recall that the stress tensor can be split into a spherical part and a deviatoric one,
$$
\tensdeux{\sigma} = \underbrace{\frac{1}{3}\trace{\tensdeux{\sigma}}\tensdeux{I}}_{\mbox{spherical}} + \underbrace{\tensdeux{S}}_{\mbox{deviatoric}}\mbox{ .}
$$
Let us introduce the spherical projection tensor $\tensquatre{S} = \frac{1}{3}\tensdeux{I}\otimes\tensdeux{I}$ and the deviatoric projection tensor $\tensquatre{D} = \tensquatre{I} - \tensquatre{S}$. In a classical model the state variable is the sand full strain $\tensdeux{\varepsilon}_s$, the internal ones contain the plastic strain $\tensdeux{\varepsilon}^p_s$, the kinematic and isotropic hardening variables $\tensdeux{\alpha}$ and $p$ \cite{bib:wood}. The free energy $\psi_s$ has the following form,
\begin{equation}
\label{eqn:psi_sable_3D}
\psi_s(\tensdeux{\varepsilon}_s,\tensdeux{\varepsilon}^p_s,\tensdeux{\alpha},p) = \frac{1}{2}\left(\tensdeux{\varepsilon}_s - \tensdeux{\varepsilon}^p_s\right):\tensquatre{K}_s\left(\tensdeux{\varepsilon}_s - \tensdeux{\varepsilon}^p_s\right) + \frac{H_k}{2}\tensdeux{\alpha}:\tensdeux{\alpha} + \frac{H_i}{2}p^2\mbox{ .}
\end{equation}
where $\tensquatre{K}_s$, $H_k$ and $H_i$ are stiffness coefficients. The state laws are directly derived from it,
\begin{equation}
\label{eqn:laws_psi_sable_3D}
\begin{array}{llrll}
\tensdeux{\sigma}^r_s & = & \frac{\partial \psi_s}{\partial \tinytensdeux{\varepsilon}_s} & = & \tensquatre{K}_s\left(\tensdeux{\varepsilon}_s - \tensdeux{\varepsilon}^p_s\right) \\
\tensdeux{A} & = & - \frac{\partial \psi_s}{\partial \tinytensdeux{\varepsilon}^p_s} & = & \tensquatre{K}_s\left(\tensdeux{\varepsilon}_s - \tensdeux{\varepsilon}^p_s\right) \\
\tensdeux{\Chi} & = & - \frac{\partial \psi_s}{\partial \tinytensdeux{\alpha}} & = & - H_k\tensdeux{\alpha} \\
R & = & - \frac{\partial \psi_s}{\partial p} & = & - H_i p \mbox{ .}\\
\end{array}
\end{equation}
To derive the complementary laws it is more convenient to define the force function $\varphi^*$ instead of the dissipation potential $\varphi$,
\begin{equation}
\label{eqn:phi_star_sable_3D}
\varphi^*_s(\tensdeux{\sigma}^{ir}_s,\tensdeux{A},\tensdeux{\Chi},R)= \textup{I}_{\{\tinytensdeux{0}\}}(\tensdeux{\sigma}^{ir}_s) + \textup{I}_{\Omega(\tinytensdeux{\tinyChi},R)}(\tensdeux{A}) \mbox{ ,}
\end{equation}
where $\Omega(\tensdeux{\Chi},R) = \left\{\tensdeux{A}\left|F(\tensdeux{A},\tensdeux{\Chi},R) \leq 0\right.\right\}$ is the elastic domain bounded by the \nomc{Drucker~--~Prager} criterion $F$ defined by \cite{bib:drucker_prager},
\begin{equation}
\label{eqn:surf_charge_sable}
F(\tensdeux{A},\tensdeux{\Chi},R) = \sqrt{J_2(\tensdeux{A} - \tensdeux{\Chi})} - \tau_y(\tensdeux{A}) - R(p) \mbox{ .}
\end{equation}
Remark that $\sqrt{J_2(.)}$ is the pseudo norm of the tensor deviatoric part implied in the plastic phenomenon. The initial threshold $\tau_y$ depends on the pressure (as it is usual in soil mechanics), on the friction coefficient $\beta$ related to the friction angle and on the cohesion parameter $C_0$, $\tau_y(\tensdeux{A}) = C_0 - \beta\,\trace{\tensdeux{A}} = \frac{\sigma_y}{\sqrt{3}}$. Since we use the dual dissipation potential, we get the complementary laws usally issued from the stress formulation (cf. table \ref{tab:resume_thermo}),
\begin{equation}
\label{eqn:norm_charge_sable}
\begin{array}{l}
\point{\tensdeux{\varepsilon}} \; \in \partial_{\tinytensdeux{\sigma}^{ir}_s}\varphi^*(\tensdeux{\sigma}^{ir}_s,\tensdeux{A},\tensdeux{\Chi},R) = \reel^{3 \times 3} \\
\begin{array}{lclcl}
\point{\tensdeux{\varepsilon}}^p_s & = & \point{\lambda}\frac{\partial F}{\partial \tinytensdeux{A}}(\tensdeux{A},\tensdeux{\Chi},R) & = & \point{\lambda}\left[\frac{\tinytensdeux{A} - \tinytensdeux{\tinyChi}}{2\sqrt{J_2(\tinytensdeux{A} - \tinytensdeux{\tinyChi})}} + \beta\tensdeux{I}\right] \\
\point{\tensdeux{\alpha}} & = & \point{\lambda}\frac{\partial F}{\partial \tinytensdeux{\tinyChi}}(\tensdeux{A},\tensdeux{\Chi},R) & = & - \point{\lambda}\frac{\tinytensdeux{A} - \tinytensdeux{\tinyChi}}{2\sqrt{J_2(\tinytensdeux{A} - \tinytensdeux{\tinyChi})}} \\
\point{p} & = & \point{\lambda}\frac{\partial F}{\partial R}(\tensdeux{A},\tensdeux{\Chi},R) & = & - \point{\lambda} \mbox{ ,}\\
\end{array} \\
\end{array}
\end{equation}
where $\point{\lambda}$ is the plastic multiplier always non negative. Its value can be found with the plastic condition $F = 0$ and the consistance condition $\point{F} = 0$,
\begin{equation}
\label{eqn:evol_charge_sable}
\left\{\begin{array}{l} F = 0 \\ \point{F} = 0 \\ \end{array}\right.\quad\Rightarrow\quad\left\{\begin{array}{l} \sqrt{J_2(\tensdeux{A} - \tensdeux{\Chi})} = \tau_y(\tensdeux{A}) + R(p) \\ \point{\lambda} = \frac{1}{H_i + \frac{H_k}{2}}\left(\frac{\tinytensdeux{A} - \tinytensdeux{\tinyChi}}{2\left(\tau_y(\tinytensdeux{A}) + R(p)\right)} + \beta\tensdeux{I}\right):\point{\tensdeux{A}} \mbox{ .}\\ \end{array}\right.
\end{equation}
Contrary to the 1D case, we cannot explicitely express a 3D dissipation potential depending on flow variables.
%==============================================================================
%
%    3.2. Unilateral wire network model
%
%==============================================================================
\subsection{Unilateral wire network model}\label{ssc:wire_3D}
According to the 1D model, we neglect the dissipation effects, and we focus on the free energy. Its stiffness cannot be reduced to the stiffness of the wire and has to account for the wire distribution in the sample, assumed to be isotropic in the following. Due to the entanglement of the wire network, it is convenient to consider continuously differentiable free energy to derive smooth relations between strain and stress at the macroscopic level. A model directly derived from the isotropic linear elasticity may be expressed in the eigen directions~; the strain and stress tensors have the same ones. Consequently the free energy is simply written using the \nomp{Lam\'e} coefficients $\lambda_w$, $\mu_w$ and the strain eigen values denoted $\varepsilon^1_w$, $\varepsilon^2_w$, $\varepsilon^3_w$ (we introduce the notations $\widetilde{\tensdeux{\varepsilon}}_w=\mbox{diag}(\varepsilon^1_w,\varepsilon^2_w,\varepsilon^3_w)$ and $\langle\widetilde{\tensdeux{\varepsilon}}_w\rangle=\mbox{diag}(\langle\varepsilon^1_w\rangle,\langle\varepsilon^2_w\rangle,\langle\varepsilon^3_w\rangle)$),
\begin{equation}
\label{eqn:psi_fil_3D}
\psi_w(\tensdeux{\varepsilon}_w) = \frac{\lambda_w}{2}\langle\varepsilon^1_w+\varepsilon^2_w+\varepsilon^3_w\rangle^2+\mu_w\left(\langle\varepsilon^1_w\rangle^2+\langle\varepsilon^2_w\rangle^2+\langle\varepsilon^3_w\rangle^2\right)\mbox{ .}
\end{equation}
The first term describes the volumic unilateral behaviour of the wire network activated by the trace of the strain. The second part concerns the shear component which is not activated in all directions simultaneously but according to the sign of the strain eigen values. The stress expressed in the eigen directions is easily derived from this previous energy,
$$
\widetilde{\tensdeux{\sigma}}^r_w = \frac{\mbox{d}\psi_w}{\mbox{d}\widetilde{\tensdeux{\varepsilon}}_w}(\widetilde{\tensdeux{\varepsilon}}_w) = \lambda_w\langle\trace{\widetilde{\tensdeux{\varepsilon}}_w}\rangle\tensdeux{I} + 2\mu_w\langle\widetilde{\tensdeux{\varepsilon}}_w\rangle\mbox{ .}
$$
In the current frame the strain stress relationship has the form,
\begin{equation}
\label{eqn:sig_fil}
\tensdeux{\sigma}^r_w = \lambda_w\langle\trace{\tensdeux{\varepsilon}_w}\rangle\tensdeux{I} + 2\mu_w\tensdeux{P}\langle\widetilde{\tensdeux{\varepsilon}}_w\rangle\tensdeux{P}^T \mbox{ ,}
\end{equation}
where $\tensdeux{P}$ depending on $\tensdeux{\varepsilon}_w$ is the passing matrix from the eigen directions to the current ones. The expression $\tensdeux{P}\langle\widetilde{\tensdeux{\varepsilon}}_w\rangle\tensdeux{P}^T$ is called the positive part of the wire strain tensor denoted $\tensdeux{\varepsilon}^{\geq}_w$. The convexity of the free energy is an open question in the general case but it is easily verified for $\mu_w = 0$ because the trace is a linear operator.
%==============================================================================
%
%    3.3. Models superposition and TexSol potentials
%
%==============================================================================
\subsection{Models superposition and \nomc{TexSol} potentials}\label{ssc:texsol_3D}
The previous model is combined according to the 1D approach. Moreover, we introduce two eventual initial stresses $\tensdeux{\sigma}^0_w$ and $\tensdeux{\sigma}^0_s$. There are generated by the deposit process under gravity which may be simulated by a discrete element software \cite{{bib:dubois_jean},{bib:moreau_bis}}. Then we can reasonably assume that eigen values of $\tensdeux{\sigma}^0_w$ are non negatives. We define the corresponding the initial strains using the elastic parts of the previous models, $\tensdeux{\varepsilon}^0_w = \tensquatre{K}^{-1}_w\tensdeux{\sigma}^0_w$ and $\tensdeux{\varepsilon}^0_s = \tensquatre{K}^{-1}_s\tensdeux{\sigma}^0_s$, where $\tensquatre{K}_w = \lambda_w\tensdeux{I}\otimes\tensdeux{I}+2\mu_w\tensquatre{I}$. The total free energy is then postulated,
\begin{equation}
\label{eqn:psi_tex_3D}
\begin{array}{rcl}
\psi(\tensdeux{\varepsilon},\tensdeux{\varepsilon}^p,\tensdeux{\alpha},p) & = & \frac{1}{2}\left(\tensdeux{\varepsilon} - \tensdeux{\varepsilon}^p + \tensdeux{\varepsilon}^0_s\right):\tensquatre{K}_s\left(\tensdeux{\varepsilon} - \tensdeux{\varepsilon}^p + \tensdeux{\varepsilon}^0_s\right) \\
 & & + \frac{\lambda_w}{2}\langle\trace{\tensdeux{\varepsilon} + \tensdeux{\varepsilon}^0_w}\rangle^2 + \mu_w\left(\tensdeux{\varepsilon} + \tensdeux{\varepsilon}^0_w\right)^{\geq}:\left(\tensdeux{\varepsilon} + \tensdeux{\varepsilon}^0_w\right)^{\geq} \\
 & & + \frac{H_k}{2}\tensdeux{\alpha}:\tensdeux{\alpha} + \frac{H_i}{2}p^2 \mbox{ .}\\
\end{array}
\end{equation}
The state laws are derived,
\begin{equation}
\label{eqn:laws_psi_tex_3D}
\begin{array}{llrll}
\tensdeux{\sigma}^r & = & \frac{\partial \psi}{\partial \tinytensdeux{\varepsilon}} & = & \tensquatre{K}_s\left(\tensdeux{\varepsilon} - \tensdeux{\varepsilon}^p\right) + \tensdeux{\sigma}^0_s + \lambda_w\langle\trace{\tensdeux{\varepsilon} + \tensdeux{\varepsilon}^0_w}\rangle\tensdeux{I} + 2\mu_w\left(\tensdeux{\varepsilon} + \tensdeux{\varepsilon}^0_w\right)^{\geq} \\
\tensdeux{A} & = & - \frac{\partial \psi}{\partial \tinytensdeux{\varepsilon}^p} & = & \tensquatre{K}_s\left(\tensdeux{\varepsilon} - \tensdeux{\varepsilon}^p \right) + \tensdeux{\sigma}^0_s \\
\tensdeux{\Chi} & = & - \frac{\partial \psi}{\partial \tinytensdeux{\alpha}} & = & - H_k\tensdeux{\alpha} \\
R & = & - \frac{\partial \psi}{\partial p} & = & - H_i p \mbox{ .}\\
\end{array}
\end{equation}
The complementary laws are derived considering the dual dissipation potential of the sand alone (cf. equation (\ref{eqn:phi_star_sable_3D})). In the simple case where $\mu_w=0$, we can complete the dual stress formulation by computing the \nomp{Legendre~--~Fenchel} transformation of the free energy $\psi$ (denoted in this case $\psi_{\circ}$) via the proposition \ref{prp:psi_conj_theo}.
\begin{equation}
\label{eqn:psi_star_tex_3D}
\psi^*_{\circ}(\tensdeux{\sigma}^r,\tensdeux{\varepsilon}^p,\tensdeux{\alpha},p) = \left\{\begin{array}{l}\scriptstyle{\frac{1}{2}\left(\tinytensdeux{\sigma}^r + \tinytensquatre{K}_s\left(\tinytensdeux{\varepsilon}^p - \tinytensdeux{\varepsilon}^0_s\right) + \tinytensdeux{\sigma}^0_w\right):\left(\tinytensquatre{K}_s + \tinytensquatre{K}^{\circ}_w\right)^{-1}\left(\tinytensdeux{\sigma}^r + \tinytensquatre{K}_s\left(\tinytensdeux{\varepsilon}^p - \tinytensdeux{\varepsilon}^0_s\right) + \tinytensdeux{\sigma}^0_w\right)} \\ \scriptstyle{-\frac{1}{2}\left(\left(\tinytensdeux{\varepsilon}^p - \tinytensdeux{\varepsilon}^0_s\right):\tinytensquatre{K}_s\left(\tinytensdeux{\varepsilon}^p - \tinytensdeux{\varepsilon}^0_s\right) + H_k\tinytensdeux{\alpha}:\tinytensdeux{\alpha} + H_ip^2\right) - \frac{1}{2}\tinytensdeux{\varepsilon}^0_w:\tinytensdeux{\sigma}^0_w} \\ \qquad\qquad\qquad\mbox{if }\trace{\tensdeux{\sigma}^r + \tensquatre{K}_s\left(\tensdeux{\varepsilon}^p - \tensdeux{\varepsilon}^0_s - \tensdeux{\varepsilon}^0_w\right)} \geq 0 \\ \scriptstyle{\frac{1}{2}\left(\tinytensdeux{\sigma}^r + \tinytensquatre{K}_s\left(\tinytensdeux{\varepsilon}^p - \tinytensdeux{\varepsilon}^0_s\right)\right):{\tinytensquatre{K}_s}^{-1}\left(\tinytensdeux{\sigma}^r + \tinytensquatre{K}_s\left(\tinytensdeux{\varepsilon}^p - \tinytensdeux{\varepsilon}^0_s\right)\right)} \\ \scriptstyle{-\frac{1}{2}\left(\left(\tinytensdeux{\varepsilon}^p - \tinytensdeux{\varepsilon}^0_s\right):\tinytensquatre{K}_s\left(\tinytensdeux{\varepsilon}^p - \tinytensdeux{\varepsilon}^0_s\right) + H_k\tinytensdeux{\alpha}:\tinytensdeux{\alpha} + H_ip^2\right)} \\ \qquad\qquad\qquad\mbox{if }\trace{\tensdeux{\sigma}^r + \tensquatre{K}_s\left(\tensdeux{\varepsilon}^p - \tensdeux{\varepsilon}^0_s - \tensdeux{\varepsilon}^0_w\right)} \leq 0 \mbox{,}\\ \end{array}\right.
\end{equation}
where $\tensquatre{K}^{\circ}_w = \lambda_w\tensdeux{I}\otimes\tensdeux{I}$. The \nomp{Legendre~--~Fenchel} transformation cannot be catched in the more general case.
%==============================================================================
%
%  4. Numerical development
%
%==============================================================================
\section{Numerical development}\label{sec:num_dev}
Starting from a coherent thermodynamical model for the \nomc{TexSol}, the next step consists in implementing it in a finite element software \cite{{bib:keryvin},{bib:kichemin_charras}}. We discuss then responses provided by the simulation of simple compression / traction tests according to the expected behaviours detailed in section \ref{sec:motivations}.
%==============================================================================
%
%    4.1. Numerical implementation
%
%==============================================================================
\subsection{Numerical implementation}\label{ssc:num_implem}
The variables being known at step $n-1$, we have to compute them at step $n$ using a predicted value of the strain increment $\Delta\tensdeux{\varepsilon}_n$. In a sake of simplicity, the initial stresses are neglected ($\tensdeux{\sigma}^0_s = \tensdeux{\sigma}^0_w = \tensdeux{0}$). Two sets of variables, ($\tensdeux{\sigma}_{s,n}, \tensdeux{\Chi}_n, p_n$) for the sand and ($\tensdeux{\sigma}_{w,n}$) for the wire network, are computed simultaneously. The stress in the wire network is directly deduced from the potential defined by (\ref{eqn:psi_fil_3D}). For the sand the relations given in (\ref{eqn:laws_psi_sable_3D}), (\ref{eqn:norm_charge_sable}) and (\ref{eqn:evol_charge_sable}) can be reduced to three equations depending on the three unknowns ($\tensdeux{\sigma}_{s,n}, \tensdeux{\Chi}_n, p_n$). This system is solved by a \nomp{Newton -- Raphson} method applied to the following residuals $\tensdeux{Q}^{\alpha}_n \mbox{ ; } \alpha = 1,2,3$.
$$
\begin{array}{rcl}
\tensdeux{Q}^1_n & = & \frac{p_n - p_{n-1}}{2\left(R_n + \tau_{y,n}\right)}\left(\tensdeux{S}_{s,n} - \tensdeux{\Chi}_n + 2\left(R_n + \tau_{y,n}\right)\beta\tensdeux{I}\right) + \Delta\tensdeux{\varepsilon}_n \\
 & & - {\tensquatre{K}_s}^{-1}\left(\tensdeux{\sigma}_{s,n} - \tensdeux{\sigma}_{s,n-1}\right) \\
\tensdeux{Q}^2_n & = & \frac{H_k\left(p_n - p_{n-1}\right)}{2\left(R_n + \tau_{y,n}\right)}\left(\tensdeux{S}_{s,n} - \tensdeux{\Chi}_n\right) + \tensdeux{\Chi}_n - \tensdeux{\Chi}_{n-1} \\
Q^3_n & = & \frac{1}{\left(R_n + \tau_{y,n}\right)\left(H_k + 2H_i\right)}\left(\tensdeux{S}_{s,n} - \tensdeux{\Chi}_n + 2\left(R_n + \tau_{y,n}\right)\beta\tensdeux{I}\right):\left(\tensdeux{\sigma}_{s,n} - \tensdeux{\sigma}_{s,n-1}\right) \\
 & & + p_n - p_{n-1} \mbox{ ,}\\
\end{array}
$$
where $\alpha = 1$ corresponds to equations (\ref{eqn:non_sliding}), (\ref{eqn:laws_psi_sable_3D})$_1$, (\ref{eqn:norm_charge_sable})$_{2,4}$ and (\ref{eqn:evol_charge_sable})$_1$, $\alpha = 2$ corresponds to equations (\ref{eqn:laws_psi_sable_3D})$_3$, (\ref{eqn:norm_charge_sable})$_{3,4}$ and (\ref{eqn:evol_charge_sable})$_1$ and finally $\alpha = 3$ corresponds to equations (\ref{eqn:norm_charge_sable})$_4$ and (\ref{eqn:evol_charge_sable})$_{1,2}$ (in all these equations, $R_n$ is calculated using the equation (\ref{eqn:laws_psi_sable_3D})$_4$). Classically, the \nomp{Taylor} development is defined as follow,
$$
\begin{array}{rcl}
\tensdeux{Q}^{\alpha}_{n,i+1} & = & \tensdeux{Q}^{\alpha}_{n,i} + {\left(\frac{\partial \tinytensdeux{Q}^{\alpha}_n}{\partial \tinytensdeux{\sigma}_{s,n}}\right)}_i\delta\tensdeux{\sigma}_{s,n,i+1} + {\left(\frac{\partial \tinytensdeux{Q}^{\alpha}_n}{\partial \tinytensdeux{\tinyChi}_n}\right)}_i\delta\tensdeux{\Chi}_{n,i+1} + {\left(\frac{\partial \tinytensdeux{Q}^{\alpha}_n}{\partial p_n}\right)}_i\delta p_{n,i+1} \mbox{ .}\\
\end{array}
$$
The analytical formulations of the tangent matrix coefficients are given,
\begin{flushleft}
$\qquad \begin{array}{rcl}
\frac{\partial \tinytensdeux{Q}^1_n}{\partial \tinytensdeux{\sigma}_{s,n}} & = & \frac{p_n - p_{n-1}}{2\left(R_n + \tau_{y,n}\right)^2}\left[\left(R_n + \tau_{y,n}\right)\tensquatre{\mathbbm{D}} + \beta\left(\tensdeux{S}_{s,n} - \tensdeux{\Chi}_n\right)\otimes\tensdeux{I}\right] - {\tensquatre{K}_s}^{-1} \\
\frac{\partial \tinytensdeux{Q}^1_n}{\partial \tinytensdeux{\tinyChi}_n} & = & - \frac{p_n - p_{n-1}}{2\left(R_n + \tau_{y,n}\right)}\tensquatre{I} \\
\frac{\partial \tinytensdeux{Q}^1_n}{\partial p_n} & = & \frac{R_n + \tau_{y,n} + H_i\left(p_n - p_{n-1}\right)}{2\left(R_n + \tau_{y,n}\right)^2}\left(\tensdeux{S}_{s,n} - \tensdeux{\Chi}_n\right) + \beta\tensdeux{I} \\
\end{array}$
\end{flushleft}
\begin{flushleft}
$\qquad \begin{array}{rcl}
\frac{\partial \tinytensdeux{Q}^2_n}{\partial \tinytensdeux{\sigma}_{s,n}} & = & \frac{H_k\left(p_n - p_{n-1}\right)}{2\left(R_n + \tau_{y,n}\right)^2}\left[\left(R_n + \tau_{y,n}\right)\tensquatre{\mathbbm{D}} + \beta\left(\tensdeux{S}_{s,n} - \tensdeux{\Chi}_n\right)\otimes\tensdeux{I}\right] \\
\frac{\partial \tinytensdeux{Q}^2_n}{\partial \tinytensdeux{\tinyChi}_n} & = & \left(1 - H_k\frac{p_n - p_{n-1}}{2\left(R_n + \tau_{y,n}\right)}\right)\tensquatre{I} \\
\frac{\partial \tinytensdeux{Q}^2_n}{\partial p_n} & = & H_k\frac{R_n + \tau_{y,n} + H_i\left(p_n - p_{n-1}\right)}{2\left(R_n + \tau_{y,n}\right)^2}\left(\tensdeux{S}_{s,n} - \tensdeux{\Chi}_n\right) \\
\end{array}$
\end{flushleft}
\begin{flushleft}
$\qquad \begin{array}{rcl}
\frac{\partial Q^3_n}{\partial \tinytensdeux{\sigma}_{s,n}} & = & C^1_n\left(2\tensdeux{S}_{s,n} - \tensdeux{S}_{s,n-1} - \tensdeux{\Chi}_n + 2\left(R_n + \tau_{y,n}\right)\beta\tensdeux{I}\right) \\
 & & + C^2_n\left(\tensdeux{S}_{s,n} - \tensdeux{\Chi}_n\right):\left(\tensdeux{\sigma}_{s,n} - \tensdeux{\sigma}_{s,n-1}\right)\tensdeux{I} \\
\frac{\partial Q^3_n}{\partial \tinytensdeux{\tinyChi}_n} & = & - C^1_n\left(\tensdeux{\sigma}_{s,n} - \tensdeux{\sigma}_{s,n-1}\right) \\
\frac{\partial Q^3_n}{\partial p_n} & = & 1 + C^3_n\left(\tensdeux{S}_{s,n} - \tensdeux{\Chi}_n - 2\left(R_n + \tau_{y,n}\right)\beta\tensdeux{I}\right):\left(\tensdeux{\sigma}_{s,n} - \tensdeux{\sigma}_{s,n-1}\right) \mbox{ ,}\\
\end{array}$
\end{flushleft}
where $C^1_n\left(R_n,\tau_{y,n}\right) = \frac{1}{\left(R_n + \tau_{y,n}\right)\left(H_k + 2H_i\right)}$, $C^2_n\left(R_n,\tau_{y,n}\right) = \frac{\beta C^1_n}{R_n + \tau_{y,n}}$ and finally $C^3_n\left(R_n,\tau_{y,n}\right) = \frac{H_i C^1_n}{R_n + \tau_{y,n}}$. The algorithm is schematized in the table \ref{tab:algo} (where $(\zeta^1_n,\zeta^2_n,\zeta^3_n)=(\tensdeux{\sigma}_{s,n},\tensdeux{\Chi}_n,p_n)$).
%\begin{table}[htbp]
%\begin{center}
%\begin{tabular}{c}
%\xymatrix{
% & & \ar@{->}[d]^-{\displaystyle{\;n=1\mbox{ and }\zeta^{\gamma}_0\mbox{ known}}} \\
% & & *+[F]{\begin{array}{c}\quad\mbox{Elastic Prediction}\quad \\ \zeta^{\gamma}_n = \mathscr{E}\left(\zeta^{\gamma}_{n-1}\right) \end{array}} \ar@(l,u)[lld] \\
%{\bullet} \ar[d]^-{F \geq 0} \ar@(rd,d)[rrr]^-{F < 0} & & & *+[F]{n=n+1} \ar@(u,r)[ul] \\
%{\bullet} \ar[rr]^-{H_i+H_k=0} \ar@(d,l)[drr]_-{H_i+H_k\not=0} & & *+[F]{\begin{array}{c}\mbox{Plastic correction by} \\ \mbox{projection} \\ \zeta^{\gamma}_n = \mathscr{C}_p\left(\zeta^{\gamma}_{n-1}\right) \end{array}} \ar@(dr,d)[ru] \\
% & & *+[F]{\begin{array}{c}\;\mbox{Plastic correction by \nomp{Newton~--~Raphson}}\; \\ \zeta^{\gamma}_{n,i+1} = \mathscr{C}_{NR}\left(\left(\frac{\partial Q^{\alpha}_n}{\partial \zeta^{\gamma}_{n}}\right)_i,\zeta^{\gamma}_{n,i},Q^{\alpha}_{n,i}\right) \end{array}} \ar@(dr,d)[ruu]}
%\end{tabular}
%\end{center}
%\caption{Solution algorithm}
%\label{tab:algo}
%\end{table}
This last one being quite complex for the sand, we have compared the results given by the previous integration law and strategy with the one developped in the \nomc{Cast3M} software where a \nomp{Drucker~--~Prager} finite element model is available. Since we got a good agreement with both implementations, we focus our attention on the coupled sand/wire model of \nomc{TexSol} involving a unilateral behaviour.
%==============================================================================
%
%    4.2. Patch test
%
%==============================================================================
\subsection{Patch test}\label{ssc:patch_test}
\begin{figure}[htbp]
\begin{center}
\includegraphics[width=11cm]{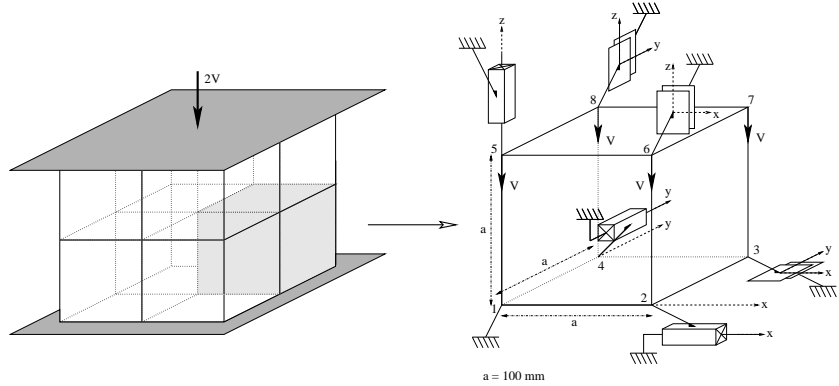}\\
\end{center}
\caption{Patch test}
\label{fig:mono_elem}
\end{figure}

In a first step, the simple patch test considered is a single Q1-\nomp{Lagrange} hexahedron finite element submitted to a traction/compression loading (cf. figure \ref{fig:mono_elem}).
More precisely, a confinement pressure is prescribed via a cohesion behaviour on the material \cite{bib:radjai_preechawuttipong_peyroux} depending on a single coefficient $C_0$. A displacement is imposed on the upperside. Four models are compared to underline the pertinency of the two unilateral behaviour laws. Two of them are considered to obtain some limit behaviours~; the first one denoted \textit{Sand}, is free of wire~; the second one denoted \textit{Reinforced sand}, is a superposition of a sand model and an elastic ``bilateral'' model of the reinforcement. The ``unilateral'' \nomc{TexSol} model referred to \S \ref{ssc:texsol_3D} is denoted \textit{Texsol}. A particular model is added denoted \textit{Spherical Texsol} corresponding to the previous one with $\mu_w=0$.
\begin{itemize}
\item \nomp{Elasticity}~: $E_s = 200000 \mbox{ MPa}$, $\nu_s = 0.4$, $E_w = 100000 \mbox{ MPa}$, $\nu_w = 0.3$
\item \nomp{Plasticity}~: $C_0 = 50 \mbox{ MPa}$, $\theta_f = 0.1$, $H_k = 100 \mbox{ MPa}$, $H_i = 100 \mbox{ MPa}$
\end{itemize}
\begin{figure}[htbp]
\begin{center}
\vspace{0.5cm}
\includegraphics[width=13cm]{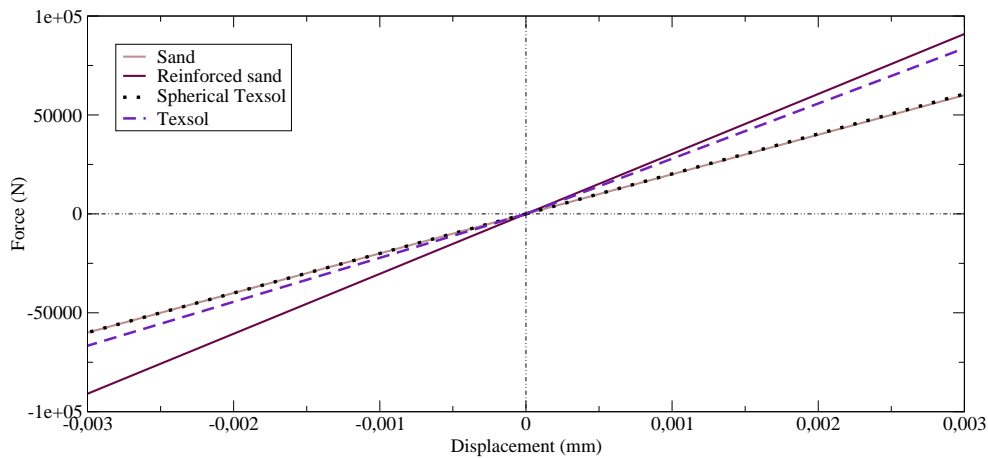}
\end{center}
\caption{Zoom on the elastic range of the models}
\label{fig:patch_elas}
\end{figure}
The \textit{Sand} and the \textit{Reinforced sand} appear clearly as two elastic bounds for \nomc{TexSol} models (cf. figure \ref{fig:patch_elas}). At this stage, the \textit{Spherical Texsol} does not differ from the \textit{Sand}. On the contrary, the \textit{Texsol} is close to the \textit{Sand} in compression and close to the \textit{Reinforced sand} in traction. For the two loadings the limit models reveal to be the upper bounds.
\begin{figure}[htbp]
\begin{center}
\vspace{0.8cm}
\includegraphics[width=13cm]{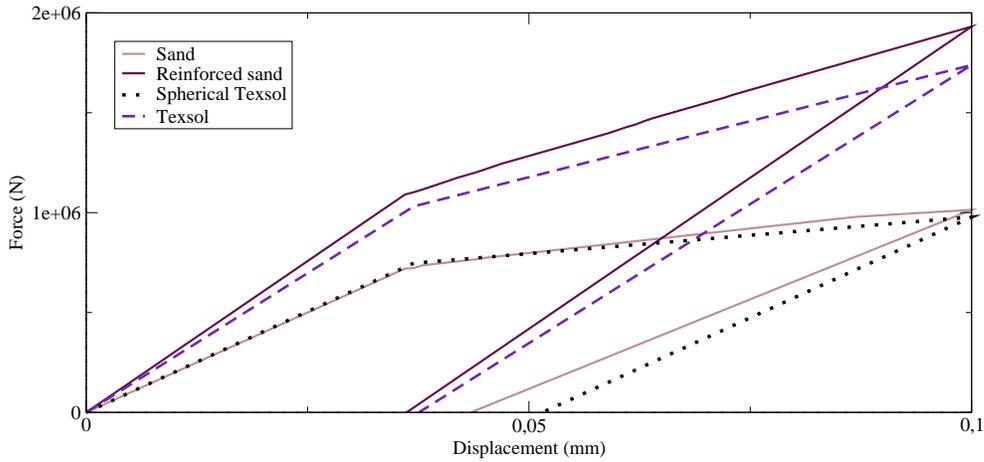}
\end{center}
\caption{Different material behaviours on a traction patch test}
\label{fig:patch_trac}
\end{figure}
\begin{figure}[htbp]
\begin{center}
\vspace{0.8cm}
\includegraphics[width=13cm]{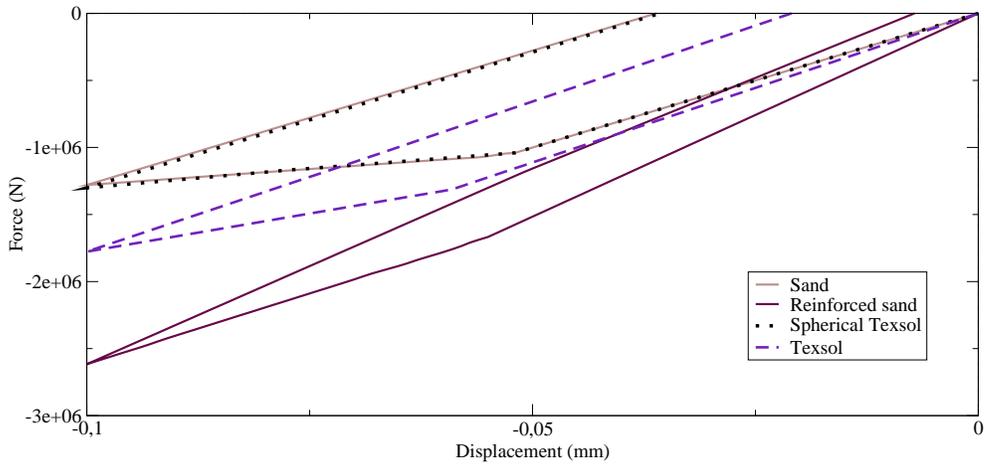}
\end{center}
\caption{Different material behaviours on a compression patch test}
\label{fig:patch_comp}
\end{figure}
For a loading-unloading traction process, the \textit{Texsol} model behaves almost like the \textit{Reinforced sand} as expected (cf. figure \ref{fig:patch_trac}). The \textit{Spherical Texsol} does not improve significantly the \textit{Sand} (cf. figure \ref{fig:patch_trac} and \ref{fig:patch_comp}). Consequently, the \textit{Spherical Texsol} does not account for the numerical results given in the figure \ref{fig:force_crushing} for the same kind of experiment - even roughly.
%==============================================================================
%
%    4.3. Cyclic loading
%
%==============================================================================
\subsection{Cyclic loading}\label{ssc:cyclic_loading}
\begin{figure}[htbp]
\begin{center}
\includegraphics[width=5cm]{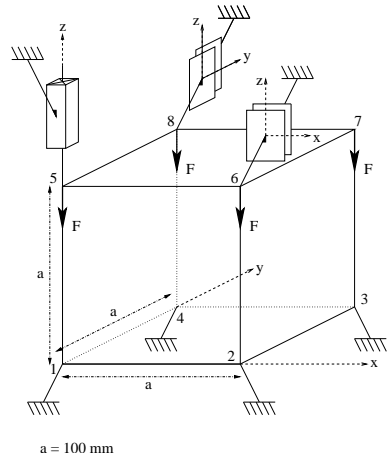}
\end{center}
\caption{Force managed traction / compression test}
\label{fig:mono_elem_bis}
\end{figure}

\begin{figure}[htbp]
\begin{center}
\vspace{0.5cm}
\includegraphics[width=13cm]{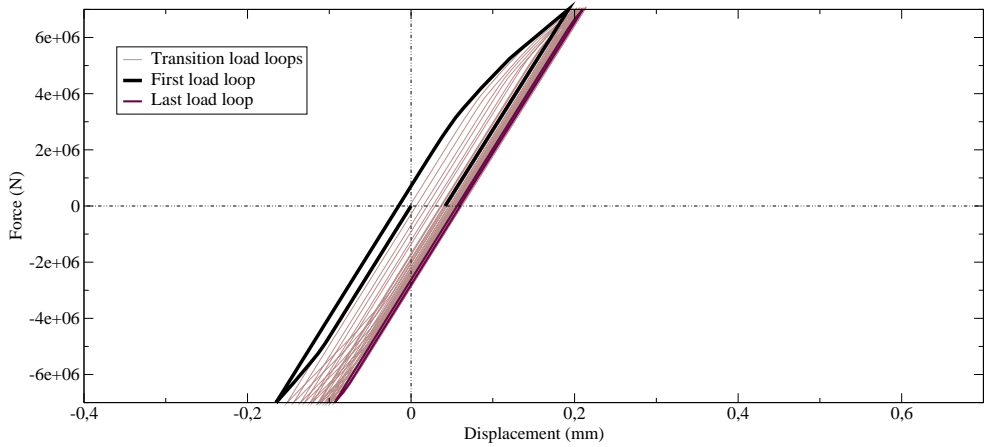}
\end{center}
\caption{Reinforced sand behaviour on a cyclic test (20 loops)}
\label{fig:patch_cycl_reinfor}
\end{figure}

\begin{figure}[htbp]
\begin{center}
\vspace{0.5cm}
\includegraphics[width=13cm]{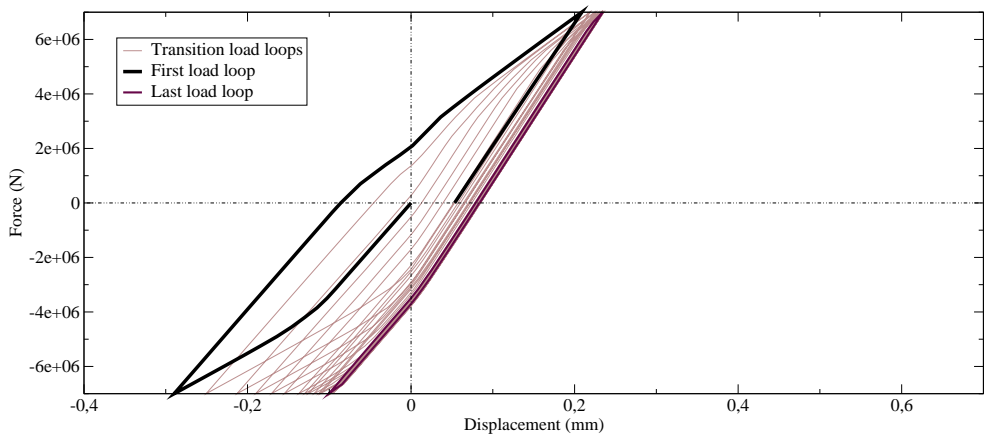}
\end{center}
\caption{Texsol behaviour on a cyclic test (20 loops)}
\label{fig:patch_cycl_texsol}
\end{figure}

\begin{figure}[htbp]
\begin{center}
\vspace{0.5cm}
\includegraphics[width=13cm]{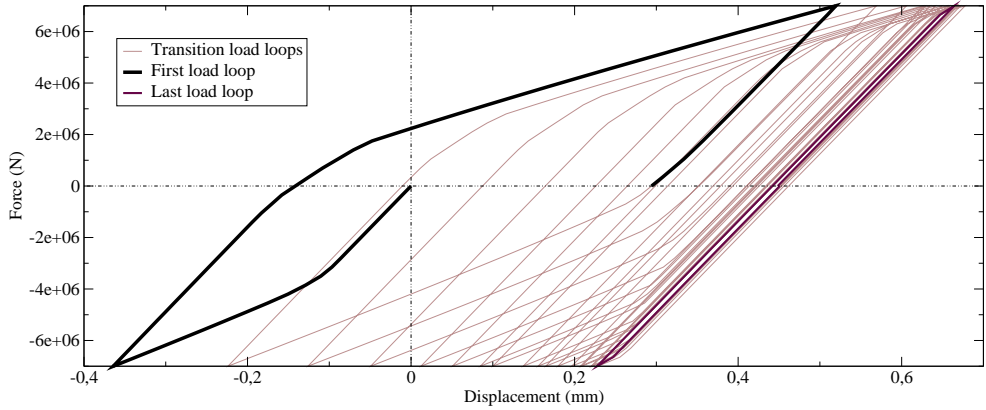}
\end{center}
\caption{Sand behaviour on a cyclic test (20 loops)}
\label{fig:patch_cycl_sand}
\end{figure}

\nomc{TexSol} embankments may be submitted to vibrating solicitations. A cyclic test based on the test represented in the figure \ref{fig:mono_elem_bis} (where the displacements are fixed on the lower side and the sollicitation managed by force) is performed to underline the contribution of the ``unilateral'' reinforcement due to the wire network.
Some material parameters are changed to apply a greater amplitude of loading on it : $C_0 = 150 \mbox{ MPa}$, $\theta_f = 0.02$, $H_k = 5 \mbox{ MPa}$, $H_i = 1 \mbox{ MPa}$. \textit{Reinforced sand}, \textit{Texsol} and \textit{Sand} are compared in the figures \ref{fig:patch_cycl_reinfor}, \ref{fig:patch_cycl_texsol} and \ref{fig:patch_cycl_sand}.
For the three models, the response tends to be stabilized after 20 cycles. But for \textit{Reinforced sand} and \textit{Texsol} the stabilization is reached before 10 loops. Moreover the residual displacement of \textit{Texsol} is $30$ percents bigger than that of \textit{Reinforced sand} and five times smaller than that of \textit{Sand}. This last result highlights the advantages of \nomc{TexSol} reinforcement. An other effect of the ``unilateral'' wire in the \textit{Texsol} model is clearly illustrated by the curvature changes when the displacement switches sign in the figure \ref{fig:patch_cycl_texsol}.
%==============================================================================
%
%    4.4. Compression test
%
%==============================================================================
\subsection{Compression test}\label{ssc:compression_test}
\begin{figure}[htbp]
\begin{center}
\includegraphics[width=13cm]{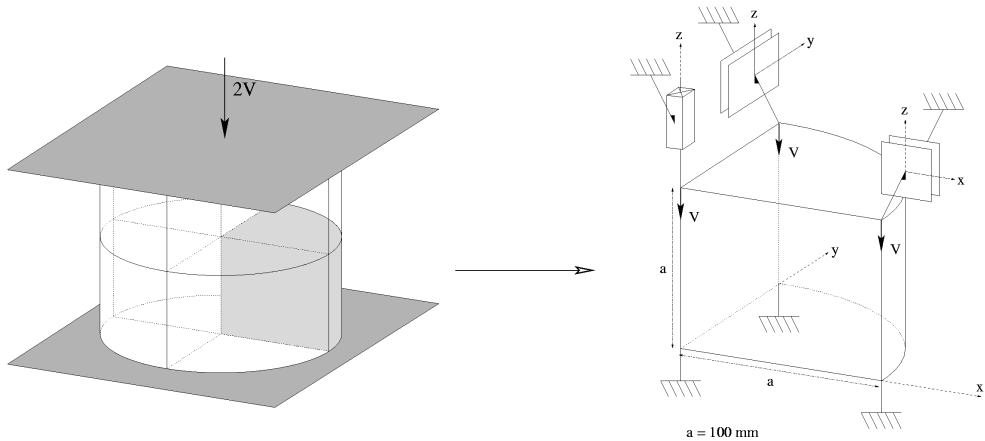}
\end{center}
\caption{Compression test}
\label{fig:cylinder}
\end{figure}

\begin{figure}[htbp]
\begin{center}
\includegraphics[width=13.9cm]{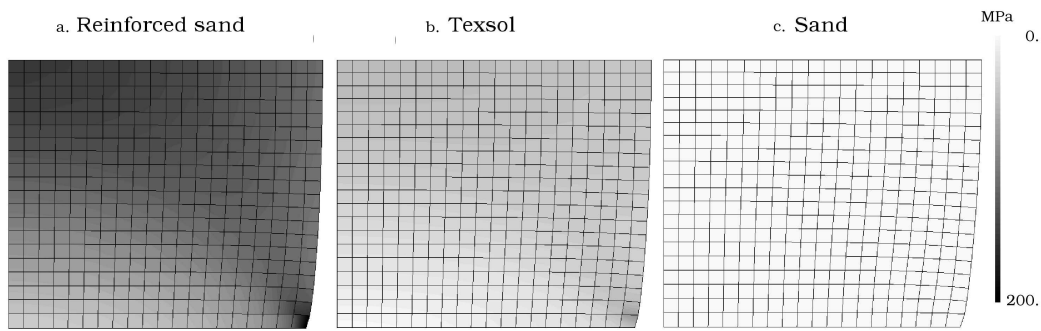}
\end{center}
\caption{Wire equivalent \nomp{Von-Mises} stress}
\label{fig:sf}
\end{figure}

\begin{figure}[htbp]
\begin{center}
\includegraphics[width=13.9cm]{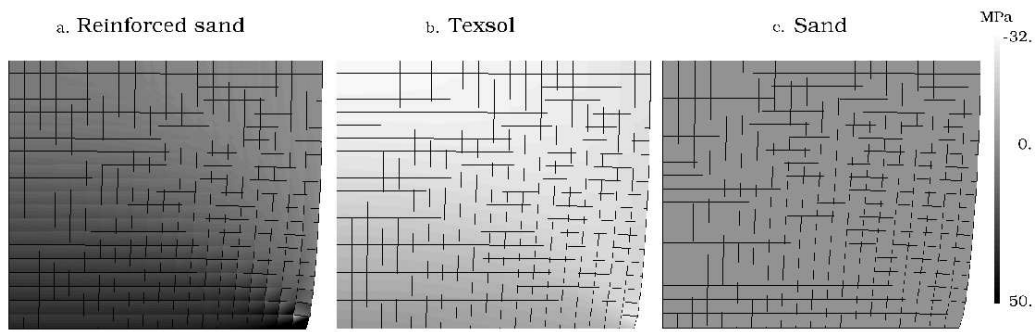}
\end{center}
\caption{Wire pressure}
\label{fig:pf}
\end{figure}

In soil mechanics it is usual to carry out a triaxial test with a prescribed confinement pressure (cf. fig \ref{fig:cylinder}). Considering the previous numerical results of the \S\ref{ssc:patch_test}, the \textit{Spherical Texsol} model is no more studied. Only the three other cases are compared in a loading compression test (the bulk mesh is described in \cite{bib:laniel}).
The contribution of the wire in \nomc{TexSol} to the mechanical strenght is illustrated by the spacial distribution of two stresses : the full stress $\tensdeux{\sigma}$ and the wire stress $\tensdeux{\sigma}_w$. The distribution of the full stress is identical in the three models with a level for \textit{Texsol} between the two others.
\begin{figure}[htbp]
\begin{center}
\includegraphics[width=13.9cm]{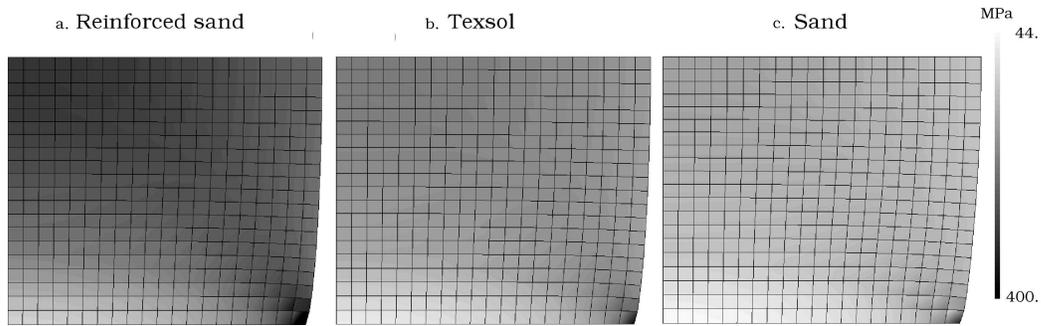}
\end{center}
\caption{Equivalent \nomp{Von-Mises} stress}
\label{fig:vm}
\end{figure}
The main part of stress is located in the center of the bulk expect a localized concentration on the right lower corner. The contribution of the wire in the stress tensor ($\tensdeux{\sigma}_w$) is split into its deviatoric part and its spherical one (pressure). Both parts are identically null for \textit{Sand} (cf. figure \ref{fig:sf}c and \ref{fig:pf}c).
The elasticity of the reinforcement is activated only in tensile directions for the \textit{Texsol} and in all directions for the \textit{Reinforced sand} ; this explains the different full stress levels in the figure \ref{fig:vm} and the different wire stress levels in the figure \ref{fig:sf}.
The nature of the reinforcement due to the wire is clearly illustrated in the figure \ref{fig:pf}. The wire pressure in the \textit{Sand} sample is of course identically null. It is negative in the \textit{Texsol} wire (traction behaviour) according to the unilaterality condition expressed in the equation (\ref{eqn:sig_fil}) whereas the pressure in the reinforcement of the \textit{Reinforced sand} is almost everywhere positive.
%==============================================================================
%
%  5. Conclusion and perspectives
%
%==============================================================================
\section{Conclusion and prospects}\label{sec:conclusion}
In this paper a coherent thermodynamical model is proposed to account for numerical experiments (because of the lack of real experiments on the \nomc{TexSol}). The keypoint is a ``unilateral'' elasticity which model the wire network. An elastic plastic model is superposed to the previous one to obtain both strain formulation and stress formulation when it is possible. Using a finite element method, we roughly validate the expected behaviour.

The main perspective of this work is the identification of the mechanical parameters of the superposed model by series of numerical experiments in progress. In a more general framework orthotropic model is generally usefull to model the wire network.

The free energy $\psi$ considered in this work is postulated and in some cases we can write the free enthalpy via the \nomp{Legendre--Fenchel} transformation. Another approach should be to postulate the free enthalpy using a form similar to (\ref{eqn:psi_fil_3D}),
$$
\mathscr{G}_w = \frac{\nu_w}{2E_w}\langle\trace{\tensdeux{\sigma}_w}\rangle^2+\frac{1-\nu_w}{2E_w}\tensdeux{\sigma}^{\geq}_w:\tensdeux{\sigma}^{\geq}_w\mbox{ .}
$$
The link between $\mathscr{G}_w$ and $\psi^*_w$ is an open question because in a three dimensional case the convexity of $\psi$ cannot be proved. \\
\vspace{1cm} \\
\textbf{Acknowledgement} \\\\
Thanks to Dr. Keryvin from the LARMAUR (Rennes) for his theoric and logistics supports.
% \appendix
% \section{}
% \label{}

\end{document}